\documentclass[onecolumn,global]{svjour}
\usepackage[T1]{fontenc}
\usepackage[latin1]{inputenc}
\usepackage{geometry}
\usepackage{float}
\usepackage{amsmath}
\usepackage{graphicx}
\usepackage{amssymb}
\usepackage[numbers]{natbib}

\makeatletter

\providecommand{\tabularnewline}{\\}

\usepackage{float}

\def\bGamma{{\mathbf\Gamma}}
\def\bGammap{{{\mathbf\Gamma}'}}

\def\bXi{{\mathbf\Xi}}

\usepackage[british]{babel}
\makeatother
\begin{document}

\title{Eddy diffusivity in convective hydromagnetic systems}

\titlerunning{Eddy diffusivity in convective hydromagnetic systems}

\author{M. Baptista \inst{1} S. M. A. Gama \inst{1} V. A. Zheligovsky \inst{2,3,4}}

\authorrunning{M. Baptista}

\institute{Centro de Matem\'{a}tica da Universidade do Porto\\
 and Departamento de Matem\'{a}tica Aplicada\\
 Faculdade de Ci\^{e}ncias da Universidade do Porto\\
 Rua do Campo Alegre, 687, 4169-007 Porto, Portugal\\
\email{mbaptist@fc.up.pt} \\
 \and International Institute of Earthquake Prediction Theory and
Mathematical Geophysics\\
 79 bldg.2, Warshavskoe ave., 117556 Moscow, Russian Federation\\
 \and Institute of Mechanics, Lomonosov Moscow State University\\
 1, Michurinsky ave., 119899 Moscow, Russian Federation\\
 \and Observatoire de la C\^{o}te d'Azur, CNRS\\
 U.M.R. 6529, BP 4229, 06304 Nice Cedex 4, France}

\maketitle

\abstract{An eigenvalue equation, for linear instability modes involving large
scales in a convective hydromagnetic system, is derived in the framework
of multiscale analysis. We consider a horizontal layer with electrically
conducting boundaries, kept at fixed temperatures and with free surface
boundary conditions for the velocity field; periodicity in horizontal
directions is assumed. The steady states must be stable to short (fast)
scale perturbations and possess symmetry about the vertical axis,
allowing instabilities involving large (slow) scales to develop. We
expand the modes and their growth rates in power series in the scale
separation parameter and obtain a hierarchy of equations, which are
solved numerically. Second order solvability condition yields a closed
equation for the leading terms of the asymptotic expansions and respective
growth rate, whose origin is in the (combined) eddy diffusivity phenomenon.
For about $10\%$ of randomly generated steady convective hydromagnetic
regimes, negative eddy diffusivity is found.}

\section{Introduction}

According to the present-day paradigm, magnetic fields of most astrophysical
objects -- the Earth and the outer planets of the Solar system having
a molten metal fluid core \citep{Merr}, the Sun \citep{Prie} and
other stars, and even entire galaxies \citep{RuSoShu} -- owe their
existence to convective hydromagnetic processes \citep{Moff,Park,Zeld}.
Convection in the presence of a magnetic field obeys a familiar set
of equations: the Navier-Stokes equation with Lorentz and Archimedes
forces for the flow, the magnetic induction equation for the magnetic
field, and the heat equation for temperature. Virtually no analytic
solutions of this system of equations are known except for some significantly
reduced cases \citep{PoZa}, under the assumption that certain symmetries
are present, or for specific initial conditions. The referred set
of equations may be used to simulate the evolution of astrophysical
convective hydromagnetic systems. This approach was followed in \citep{JoRo00,RoJo,JoRo},
where magneto-convection in an idealised plane layer was considered,
and in \citep{GlRo95a,GlRo95b,GlRo96a,GlRo96b,GlRo97a,GlRo97b,RoGl,Jones},
where the outer core of the Earth was modelled by three-dimensional
equations of hydromagnetic convection in a spherical layer and, as
a result, the predominant dipole morphology of the Earth's magnetic
field was correctly reproduced in computations.

However, in practice, accurate simulations for geo- and astrophysical
real parameter values are close to impossible, because the limited
power of available computers prohibits computations with the adequate
spatial (insufficient memory) and temporal (insufficient CPU power)
resolution. The simulations done by Glatzmaier and Roberts \citep{GlRo95a,GlRo95b,GlRo96a,GlRo96b,GlRo97a,GlRo97b}
were performed for parameter values differing by several orders of
magnitude from those characterising the outer liquid core of the Earth.
Nevertheless, the agreement between these simulations and the geodynamo
is surprising \citep{Jones}. The existence of sharp contrast spatial
structures (e.g., Ekman boundary layer, which emerges in convective
flows in a rotating layer with no-slip boundary conditions, and the
instability of which may be a source of dynamo \citep{PoGiSo01a,PoGiSo01b,PoGiSo03,Rotv,Stell})
and the prominent role which turbulence plays in the generation of
magnetic fields, indicate that such a high resolution is indeed necessary
in simulations. Core-mantle coupling, which is believed to cause decade
variations of the length of day, is another geophysical phenomenon
involving small scales (topographic features at the boundary are unlikely
to exceed 5 km in amplitude; see \citep{Merr}). Validity of numerical
techniques for smoothening the resolution cut-off, such as employment
of hyperviscosity, is questionable \citep{ZhaJo,Sars,ZhaSchu,Bus}.

A significant uncertainty in rheology relations \citep{ChrPel} and
parameter values \citep{PelPel} in the convective hydromagnetic equations
(for instance, estimates of thermal diffusivity for the Earth core
differ in several orders of magnitude \citep{Merr}), makes it desirable
to investigate typical regimes of behaviour of the solutions by varying
the parameters in certain ranges and finding locations, in the parameter
space, of bifurcations marking drastic changes in behaviour \citep{ChrOlGl}.
A purely numerical approach to the implementation of this task would
require a further significant increase of the amount of computations.
Consequently, a semi-analytic approach, employing asymptotic relations,
is unavoidable as an alternative to both purely analytic and numerical
approaches. Here we employ it to consider the problem of linear stability
of three-dimensional convective hydromagnetic steady states in a layer.

The characteristic spatial scale of the perturbed steady state is
supposed to be much larger than that of the steady state. The ratio
of the spatial scale of the flow (fast spatial variable) to the large
scale of perturbation (slow variable), $\varepsilon$, is a small
parameter. (By small- and large-scale vector fields we refer to fields
involving spatial scales of the order of the width of the layer and
much larger scales, respectively.) Applying methods of the general
theory of homogenisation of differential operators \citep{Bens,Ole,Jik,Cio},
we expand perturbation modes and their growth rates in asymptotic
series in the parameter $\varepsilon$ and obtain a homogenised operator
in slow variables, acting on mean fields. Eigenvalues of this operator
control stability to large-scale perturbations. The advantage of this
approach stems from opening the possibility to disentangle the large
and small scales and fully resolve small scales by solving the so-called
auxiliary problems.

Generically, the multiscale analysis reveals the presence of $\alpha$-effect
(see \citep{FrZhSu,SuShe,Zhe91}). The homogenised linear operator
is then the first-order differential operator. Consequently, the system
is generically unstable, since the spectrum of the operator is symmetric
about the origin (if a mode $\mathbf{W}(x)$ is associated with an
eigenvalue $\lambda$, then $\mathbf{W}(-x)$ is a mode associated
with the eigenvalue $-\lambda$). In convective hydromagnetic systems
which possess symmetry about a vertical axis or parity-invariance,
$\alpha$-effect is not present and the homogenised equations involve
a second-order partial differential operator, whose eigenfunctions
are Fourier harmonics. Its eigenvalues may be positive, implying instability.
This phenomenon is referred to as negative (combined) eddy diffusivity
\citep{Star}. Instability of this kind is weak: in the presence of
$\alpha$-effect the growth rate of the dominant perturbations is
$O(\varepsilon)$, whereas it is $O(\varepsilon^{2})$ when $\alpha$-effect
is absent. Evaluation of eddy tensors emerging in the homogenised
equations requires solution of auxiliary problems, which are linear
elliptic partial differential equations in fast variables. With just
a single characteristic spatial scale involved, they are not too demanding
numerically.

Multiscale asymptotic analysis was successfully applied to various
problems of hydrodynamics and magnetohydrodynamics. The effect of
negative eddy viscosity arises in two-dimensional \citep{SiYa,SiFre,GaVeFr}
and three-dimensional \citep{Yu90,DubFr,WiGaFr} hydrodynamic systems,
if the flow is parity-invariant or if it is a Beltrami field (in \citep{Yu90}
large scale along only one direction was assumed). Eddy diffusivity
can be complex \citep{Wirth,WiGaFr}. In generic hydrodynamic systems,
which do not possess the properties mentioned above, similar expansions
indicate the presence of the so-called AKA-effect (i.e. anisotropic
kinetic $\alpha$-effect) \citep{FrZhSu,SuShe}. In passive scalar
transport systems, eddy diffusivity can only enhance molecular diffusivity
\citep{Bife,VeAv}.

In the kinematic dynamo problem (concerning magnetic field generation,
when the feedback influence of magnetic field on the flow via the
Lorentz force is neglected), multiscale expansions were apparently
first introduced in \citep{Chi} and \citep{Sow} (where scale separation
was related to fast rotation of the layer of conducting fluid). Similar
asymptotic expansions in the kinematic problem (for flows, the amplitude
of which may depend on the scale ratio) predict occurrence of $\alpha$-effect
\citep{Vi86,Vi87,Zhe91}. Generation of large-scale magnetic field
by the negative magnetic eddy diffusivity mechanism is possible for
parity-invariant steady \citep{Lano,Zhe01,Yu01} (in \citep{Yu01},
large scale along only one direction was assumed) or time-periodic
flows \citep{ZhePo}, and by convective Bisshopp cell patterns \citep{Zhe05},
symmetric about the vertical axis. Combined eddy diffusivity tensors
for large-scale perturbations of both the flow and magnetic field
constituting a parity-invariant three-dimensional MHD steady state
were derived in \citep{Zhe03}.

In the papers cited above two different scales were present in the
system. Multiscale expansions with three spatial scales were employed
in \citep{Po05,Po05mjg} to study the small-angle instability \citep{Cox}
in convection in a rotating layer.

Evolution of a mean hydrodynamic large-scale perturbation in the weakly
nonlinear regime was considered in the absence of magnetic field for
two-dimensional parity-invariant space-periodic flows \citep{GaVeFr,PoPaSu},
and for three-dimensional MHD systems \citep{Zhe05fz}. In \citep{Zhe05fz}
it is not required that the MHD state, nonlinear stability of which
is examined, is either space periodic or steady; equations for the
mean flow and magnetic field are generalisations of the Navier-Stokes
and magnetic induction equation with an anisotropic (in general) combined
eddy diffusivity tensor and quadratic eddy advection.

In section \ref{sec:Equations-describing-convection}, we present
the equations for thermal convection in the presence of magnetic field
and discuss boundary conditions and symmetries. In section \ref{sec:Linearised large-scale CHM equation},
the multiscale formalism is applied. In section \ref{sub:The-two-scales-expansion},
an eigenfunction of the linearisation of a convective hydromagnetic
system and its associated eigenvalue are expanded in a power series
of the scale separation parameter and a hierarchy of equations is
derived. In section \ref{sub:Solvability-Conditions}, the solvability
condition, which plays an important role in solution of equations
of this hierarchy and in derivation of an equation for the mean part
of perturbation in the leading order, is discussed. In sections \ref{sub:Equations-at-order-0}
and \ref{sub:Equations-at-order-1}, the first two (order 0 and order
1) equations in the hierarchy are expressed as a linear combination
of the so-called auxiliary problems. In section \ref{sub:The-mean-field-equations},
we consider the solvability condition for equations at order 2 and
thereby derive the eigenvalue equation for the mean part of the leading
terms in the expansions of the instability modes and their growth
rates. At this stage emerges the homogenised combined eddy diffusivity
operator acting on mean fields. In section \ref{sec:Numerical-Results},
we briefly describe the numerical procedure for solving the auxiliary
problems and present a set of basic fields which lead to large-scale
instability for appropriate physical parameters (namely molecular
diffusivities). Finally (section \ref{sec:Concluding-Remarks}), we
comment on possible extensions and limitations of the application
of multiscale techniques employed here to study the instability of
convective flows in the presence of magnetic field.

\section{Equations of thermal convection in the presence of magnetic field\label{sec:Equations-describing-convection}}

\subsection{Time evolution of a convective hydromagnetic system}

Magnetic field generation by thermal convection is governed by the
Navier-Stokes equation, the magnetic induction equation and heat transfer
equation \citep{Chan}: \begin{eqnarray}
\partial_{t}\mathbf{V} & = & \mathbf{V}\times(\partial\times\mathbf{V})-\partial p+\nu\partial^{2}\mathbf{V}\nonumber \\
 &  & -\mathbf{H}\times(\partial\times\mathbf{H})-\alpha(T-T_{0})\mathbf{G}+\tilde{\mathbf{F}},\nonumber \\
\partial\cdot\mathbf{V} & = & 0,\nonumber \\
\partial_{t}\mathbf{H} & = & \partial\times(\mathbf{V}\times\mathbf{H})+\eta\partial^{2}\mathbf{H}+\tilde{\mathbf{R}},\label{eq:thermoconveqs}\\
\partial\cdot\mathbf{H} & = & 0,\nonumber \\
\partial_{t}\, T & = & -(\mathbf{V}\cdot\partial)T+k\partial^{2}T+\frac{\sigma}{2}\left|\partial\times\mathbf{H}\right|^{2}+\tilde{S},\nonumber \end{eqnarray}
 where $\mathbf{V}=(V_{1},V_{2},V_{3})$, $\mathbf{H}=(H_{1},H_{2},H_{3})$
and $T$, depending on position in space $\mathbf{x}=(x_{1},x_{2},x_{3})$
and time $t$, are the velocity field, the magnetic field and the
temperature, respectively. We use the notation $\partial_{t}\equiv\partial/\partial t$,
$\partial_{i}\equiv\partial/\partial x_{i}$ and $\partial\equiv\sum_{i=1}^{3}\mathbf{e}_{i}\partial_{i}$,
where $\mathbf{e}_{i}$ is the \emph{i}th canonical vector. The term
involving $\mathbf{G}$ (gravity, $\mathbf{G}=-g\,\mathbf{e}_{3}$
for a horizontal layer) is the buoyancy force due to temperature variation
and $\mathbf{H}\times(\partial\times\mathbf{H})$ is the Lorentz force.
$\tilde{\mathbf{F}}$ represents any other body forces acting on the
fluid, $\tilde{\mathbf{R}}$ is due to imposed external currents or
magnetic fields, and $\tilde{S}$ describes the distribution of external
heat sources. $\nu$ is the kinematic viscosity, $\eta$ the magnetic
diffusivity, and $\alpha$, $k$ and $\sigma$ are parameters related
to thermal expansion, thermal conductivity and electrical conductivity,
respectively. Solenoidality of magnetic field follows from the Maxwell
equations. Flows are deemed incompressible in line with the Boussinesq
approximation. Henceforth, the system of equations \eqref{eq:thermoconveqs}
will be referred to as CHM (convective hydromagnetic).

We consider the CHM equations in the spatial domain ${\cal D}=[0,L_{1}]\times[0,L_{2}]\times[0,L_{3}]$,
assuming periodicity in $x_{1}$ and $x_{2}$ directions, and considering
a finite layer in the $x_{3}$ direction. The boundary conditions
at the surface of the layer are

$\bullet$ for the velocity field: \begin{eqnarray*}
 &  & V_{3}|_{x_{3}=0,L_{3}}=0,\\
 &  & \partial_{3}V_{1}|_{x_{3}=0,L_{3}}=\partial_{3}V_{2}|_{x_{3}=0,L_{3}}=0;\end{eqnarray*}
 $\bullet$ for the magnetic field: \begin{eqnarray*}
 &  & H_{3}|_{x_{3}=0,L_{3}}=0,\\
 &  & \partial_{3}H_{1}|_{x_{3}=0,L_{3}}=\partial_{3}H_{2}|_{x_{3}=0,L_{3}}=0;\end{eqnarray*}
 $\bullet$ for the temperature field: \begin{eqnarray*}
 &  & T|_{x_{3}=0}=T_{1},\\
 &  & T|_{x_{3}=L_{3}}=T_{2}.\end{eqnarray*}
 It is convenient to introduce the variable \[
\theta=T-T_{1}-\delta Tx_{3},\]
 where $\delta T=(T_{2}-T_{1})/L_{3}$, satisfying the uniform boundary
conditions \begin{eqnarray*}
 &  & \theta|_{x_{3}=0,L_{3}}=0.\end{eqnarray*}

A solution to the CHM system may possess symmetry about the vertical
axis, provided the forcing terms $\tilde{\mathbf{F}},\tilde{\mathbf{R}},\tilde{S}$
possess the same symmetry. A vector field $\mathbf{Q}$ is called
symmetric (about the vertical axis) if \begin{eqnarray*}
 &  & Q_{1}(-x_{1},-x_{2},x_{3})=-Q_{1}(x_{1},x_{2},x_{3}),\\
 &  & Q_{2}(-x_{1},-x_{2},x_{3})=-Q_{2}(x_{1},x_{2},x_{3}),\\
 &  & Q_{3}(-x_{1},-x_{2},x_{3})=Q_{3}(x_{1},x_{2},x_{3});\end{eqnarray*}
 and anti-symmetric if \begin{eqnarray*}
 &  & Q_{1}(-x_{1},-x_{2},x_{3})=Q_{1}(x_{1},x_{2},x_{3}),\\
 &  & Q_{2}(-x_{1},-x_{2},x_{3})=Q_{2}(x_{1},x_{2},x_{3}),\\
 &  & Q_{3}(-x_{1},-x_{2},x_{3})=-Q_{3}(x_{1},x_{2},x_{3}).\end{eqnarray*}
 We call a scalar field $f$ symmetric if \[
f(-x_{1},-x_{2},x_{3})=f(x_{1},x_{2},x_{3}),\]
 and anti-symmetric if \[
f(-x_{1},-x_{2},x_{3})=-f(x_{1},x_{2},x_{3}).\]

Symmetries are essential to eliminate first order (alpha) effects.
In \citep{DubFr} parity-invariance is used to this purpose, but,
for a horizontal layer, symmetry about the vertical axis is more realistic.
Furthermore, parity invariance is inconsistent with the basic equations
\eqref{eq:thermoconveqs} for $\sigma\ne0.$ Under an appropriate
forcing, any hydromagnetic convective system will possess steady states
with these symmetries.

\subsection{Linearised CHM operator\label{sub:Linearised-CHM-operator}}

Let us consider a steady state solution, $\tilde{p}$, $\tilde{\mathbf{V}}$,
$\tilde{\mathbf{H}}$ and $\tilde{\theta}$, of the CHM system \eqref{eq:thermoconveqs}
and a small perturbation, $pe^{\lambda t}$, $\mathbf{V}e^{\lambda t}$,
$\mathbf{H}e^{\lambda t}$ and $\theta e^{\lambda t}$, of this steady
state, where $p$, $\mathbf{V}$, $\mathbf{H}$ and $\theta$ depend
only on spatial variables. In what follows, we will call the spatial
profiles of the perturbation fields, $p$, $\mathbf{V}$, $\mathbf{H}$
and $\theta$, a perturbation. Replacing $p$, $\mathbf{V}$, $\mathbf{H}$,
$\theta$, respectively, by $\tilde{p}+pe^{\lambda t}$, $\tilde{\mathbf{V}}+\mathbf{V}e^{\lambda t}$,
$\tilde{\mathbf{H}}+\mathbf{H}e^{\lambda t}$, $\tilde{\theta}+\theta e^{\lambda t}$
in \eqref{eq:thermoconveqs} and neglecting second order terms in
the perturbation, we obtain an eigenvalue problem for the perturbation:\begin{equation}
\begin{array}{l}
\mathbf{A}\mathbf{W}=\lambda\mathbf{W}+\left[\begin{array}{c}
\partial p\\
\mathbf{0}\\
0\end{array}\right],\\
\partial\cdot\mathbf{V}=0,\\
\partial\cdot\mathbf{H}=0.\end{array}\label{eq:lineqs_matrix}\end{equation}
Here, the block notation introduced in \citep{DubFr} is used:\begin{equation}
\mathbf{W}=\left[\begin{array}{c}
\mathbf{V}\\
\mathbf{H}\\
\theta\end{array}\right],\label{eq:bigW}\end{equation}
 is the $(3+3+1)$-dimensional block column vector combining the 3
components of the flow, the 3 components of the magnetic field and
the temperature field. In what follows, $(3+3+1)$-dimensional vectors
of a similar structure, will be used. The operator $\mathbf{A}$ is
obtained by linearisation of the CHM equations in the vicinity of
the steady state $\tilde{p}$, $\tilde{\mathbf{V}}$, $\tilde{\mathbf{H}}$
and $\tilde{\theta}$; it can be represented as a block matrix (acting
on $(3+3+1)$-dimensional vectors of the structure similar to \eqref{eq:bigW}):

\begin{eqnarray}
\mathbf{A} & = & \left[\begin{array}{lllll}
\nu\partial^{2}+\tilde{\mathbf{V}}\times(\partial\times\bullet)-(\partial\times\tilde{\mathbf{V}})\times &  & -\tilde{\mathbf{H}}\times(\partial\times\bullet)+(\partial\times\tilde{\mathbf{H}})\times &  & -\alpha\,\mathbf{G}\\
-\partial\times(\tilde{\mathbf{H}}\times\bullet) &  & \eta\partial^{2}+\partial\times(\tilde{\mathbf{V}}\times\bullet) &  & \mathbf{0}\\
-(\bullet\cdot\partial)\tilde{\theta}-\delta T\mathbf{e}_{3}\cdot &  & \sigma(\partial\times\tilde{\mathbf{H}})\cdot(\partial\times\bullet) &  & k\,\partial^{2}-\tilde{\mathbf{V}}\cdot\partial\end{array}\right].\label{eq:A}\end{eqnarray}
 Note that $\mathbf{A}$ preserves the symmetry of fields $\mathbf{W}$,
symmetric (or anti-symmetric) about the vertical axis.

The complete formulation of the eigenvalue problem \eqref{eq:lineqs_matrix}
involves specifying spatial periods of perturbations, which can be
any integer multiples of the periods $L_{1}$ and $L_{2}$. If the
smallest of the periodicity boxes is considered, the system of equations
\eqref{eq:lineqs_matrix} is referred to as the problem of linear
stability to short-scale perturbations.

\section{Linearised large-scale CHM equation\label{sec:Linearised large-scale CHM equation}}

\subsection{The two-scales expansion\label{sub:The-two-scales-expansion}}

In this section we construct a homogenisation of the linearised CHM
operator. Eigenvalues of the homogenised operator control linear stability
of the CHM steady state to perturbations with spatial periods large
enough for the asymptotic behaviour to set in.

Following the method applied in previous studies \citep{Lano,Zhe01,Zhe03,ZhePo},
we consider fast variables, $\mathbf{x}$, representing the short
scale dynamics, and slow variables, $\mathbf{X}=(X_{1},X_{2})=(\varepsilon x_{1},\varepsilon x_{2})$,
representing the large scale dynamics. (In a layer of finite width
only slow variables in horizontal directions are geometrically consistent.)
The parameter $\varepsilon$ is the scale separation. The perturbations
$p$, $\mathbf{V}$, $\mathbf{H}$ and $\theta$ are assumed to depend
on both fast and slow variables. We denote $\nabla_{i}\equiv\partial/\partial X_{i}$
and $\nabla\equiv\sum_{i=1}^{2}\mathbf{e}_{i}\nabla_{i}$.

Now we perform differentiation in \eqref{eq:lineqs_matrix} applying
the chain rule: $\partial_{i}\rightarrow\partial_{i}+\varepsilon\nabla_{i}$
for $i=1,2$. This implies \[
\mathbf{A}=\mathbf{A}^{(0)}+\varepsilon\mathbf{A}^{(1)}+\varepsilon^{2}\mathbf{A}^{(2)},\]
 where \begin{eqnarray}
\mathbf{A}^{(0)} & = & \left[\begin{array}{lclll}
\nu\partial^{2}+\tilde{\mathbf{V}}\times(\partial\times\bullet)-(\partial\times\tilde{\mathbf{V}})\times &  & -\tilde{\mathbf{H}}\times(\partial\times\bullet)+(\partial\times\tilde{\mathbf{H}})\times &  & -\alpha\,\mathbf{G}\\
-\partial\times(\tilde{\mathbf{H}}\times\bullet) &  & \eta\partial^{2}+\partial\times(\tilde{\mathbf{V}}\times\bullet) &  & \mathbf{0}\\
-(\bullet\cdot\partial)\tilde{\theta}-\delta T\mathbf{e}_{3}\cdot &  & \sigma(\partial\times\tilde{\mathbf{H}})\cdot(\partial\times\bullet) &  & k\,\partial^{2}-\tilde{\mathbf{V}}\cdot\partial\end{array}\right],\label{eq:a0}\\
\nonumber \\\mathbf{A}^{(1)} & = & \left[\begin{array}{lclcl}
2\nu\partial\cdot\nabla+\tilde{\mathbf{V}}\times(\nabla\times\bullet) &  & -\tilde{\mathbf{H}}\times(\nabla\times\bullet) &  & \mathbf{0}\\
-\nabla\times(\tilde{\mathbf{H}}\times\bullet) &  & 2\eta\partial\cdot\nabla+\nabla\times(\tilde{\mathbf{V}}\times\bullet) &  & \mathbf{0}\\
0 &  & \sigma(\partial\times\tilde{\mathbf{H}})\cdot(\nabla\times\bullet) &  & 2k\partial\cdot\nabla-\tilde{\mathbf{V}}\cdot\nabla\end{array}\right],\label{eq:a1}\\
\nonumber \\\mathbf{A}^{(2)} & = & \bXi\nabla^{2},\textrm{ with }\ \bXi=\left[\begin{array}{ccc}
\nu & \mathbf{0} & \mathbf{0}\\
\mathbf{0} & \eta & \mathbf{0}\\
0 & 0 & k\end{array}\right]\label{eq:a2}\end{eqnarray}
 ($\bXi$ is the molecular diffusivity tensor). Note that $\mathbf{A}^{(0)}$
and $\mathbf{A}^{(2)}$ preserve the symmetries of both symmetric
and anti-symmetric fields, but $\mathbf{A}^{(1)}$ exchanges their
symmetry.

Next we expand $\mathbf{W}$, $p$ and $\lambda$ in power series
of $\varepsilon$: \begin{equation}
\mathbf{W}=\sum_{i=0}^{n}\varepsilon^{i}\mathbf{W}^{(i)}+O(\varepsilon^{n+1}),\label{eq:Wse}\end{equation}
 \begin{equation}
p=\sum_{i=0}^{n}\varepsilon^{i}p^{(i)}+O(\varepsilon^{n+1}),\label{eq:pse}\end{equation}
 \begin{equation}
\lambda=\sum_{i=0}^{n}\varepsilon^{i}\lambda_{i}+O(\varepsilon^{n+1}).\label{eq:Lambdase}\end{equation}
 Each coefficient $\mathbf{W}^{(i)}$ and $p^{(i)}$ in the expansions
is a function of both $\mathbf{x}$ and $\mathbf{X}$. Substituting
these series in \eqref{eq:lineqs_matrix} and equating the terms in
$\varepsilon^{n}$, at each order $n$, we obtain a hierarchy of equations:

\begin{itemize}
\item order 0:\begin{eqnarray}
\mathbf{A}^{(0)}\mathbf{W}^{(0)} & = & \lambda_{0}\mathbf{W}^{(0)}+\left[\begin{array}{c}
\partial p^{(0)}\\
\mathbf{0}\\
0\end{array}\right],\label{eq:order0}\\
\partial\cdot\mathbf{V}^{(0)} & = & 0,\label{eq:order0sv}\\
\partial\cdot\mathbf{B}^{(0)} & = & 0,\label{eq:order0sh}\end{eqnarray}

\item order 1:\begin{eqnarray}
\mathbf{A}^{(0)}\mathbf{W}^{(1)} & = & -\mathbf{A}^{(1)}\mathbf{W}^{(0)}+\lambda_{0}\mathbf{W}^{(1)}+\lambda_{1}\mathbf{W}^{(0)}+\left[\begin{array}{c}
\partial p^{(1)}+\nabla p^{(0)}\\
\mathbf{0}\\
0\end{array}\right],\label{eq:order1}\\
\partial\cdot\mathbf{V}^{(1)} & = & -\nabla\cdot\mathbf{V}^{(0)},\label{eq:order1sv}\\
\partial\cdot\mathbf{B}^{(1)} & = & -\nabla\cdot\mathbf{B}^{(0)},\label{eq:order1sh}\end{eqnarray}
\newpage{}
\item order 2:\begin{eqnarray}
\mathbf{A}^{(0)}\mathbf{W}^{(2)} & = & -\mathbf{A}^{(1)}\mathbf{W}^{(1)}-\mathbf{A}^{(2)}\mathbf{W}^{(0)}+\lambda_{0}\mathbf{W}^{(2)}+\lambda_{1}\mathbf{W}^{(1)}+\lambda_{2}\mathbf{W}^{(0)}+\left[\begin{array}{c}
\partial p^{(2)}+\nabla p^{(1)}\\
\mathbf{0}\\
0\end{array}\right],\label{eq:order2}\\
\partial\cdot\mathbf{V}^{(2)} & = & -\nabla\cdot\mathbf{V}^{(1)},\label{eq:order2sv}\\
\partial\cdot\mathbf{B}^{(2)} & = & -\nabla\cdot\mathbf{B}^{(1)},\label{eq:order2sh}\end{eqnarray}

\item order $n$:\begin{eqnarray}
\mathbf{A}^{(0)}\mathbf{W}^{(n)} & = & -\mathbf{A}^{(1)}\mathbf{W}^{(n-1)}-\mathbf{A}^{(2)}\mathbf{W}^{(n-2)}+\sum_{i=0}^{n}\lambda_{i}\mathbf{W}^{(n-i)}+\left[\begin{array}{c}
\partial p^{(n)}+\nabla p^{(n-1)}\\
\mathbf{0}\\
0\end{array}\right],\label{eq:ordern}\\
\partial\cdot\mathbf{V}^{(n)} & = & -\nabla\cdot\mathbf{V}^{(n-1)},\label{eq:ordernsv}\\
\partial\cdot\mathbf{B}^{(n)} & = & -\nabla\cdot\mathbf{B}^{(n-1)}.\label{eq:ordernsh}\end{eqnarray}

\end{itemize}
 Let $\langle\bullet\rangle=(L_{1}L_{2}L_{3})^{-1}\int_{{\cal D}}\bullet dx_{1}dx_{2}dx_{3}$
denote the mean (over the fast variables in ${\cal D}$) and $\{\bullet\}=\bullet-\langle\bullet\rangle$
denote the fluctuating part of a vector or scalar field here denoted
indistinguishably by $\bullet$. The average $\langle\bullet\rangle$
is the large scale component of the respective field. It is possible
to solve recursively all equations in the hierarchy, finding all terms
of the expansions \eqref{eq:Wse}-\eqref{eq:Lambdase}. Only the equations
up to order 2 are required to derive a homogenised eigenvalue equation
for the mean parts, $\langle\mathbf{W}^{(0)}\rangle$ and $\langle p^{(0)}\rangle$,
of the leading terms. It emerges as the solvability condition for
the equation in the fast variables at order 2.

~

\subsection{Solvability Conditions\label{sub:Solvability-Conditions}}

Let $\mathbf{P}$ be the projection into the subspace of $(3+3+1)$-dimensional
fields, in which the 3-dimensional vector components are solenoidal:
\[
\mathbf{P}\left[\begin{array}{l}
\mathbf{Q}^{\mathbf{V}}\\
\mathbf{Q}^{\mathbf{H}}\\
Q^{\theta}\end{array}\right]=\left[\begin{array}{l}
\mathbf{Q}^{\mathbf{V}}-\partial Q^{\mathbf{V}p}\\
\mathbf{Q}^{\mathbf{H}}-\partial Q^{\mathbf{H}p}\\
Q^{\theta}\end{array}\right],\]
 where $\partial^{2}Q^{\mathbf{V}p}=\partial\cdot\mathbf{Q}^{V}$
and $\partial^{2}Q^{\mathbf{H}p}=\partial\cdot\mathbf{Q}^{H}$. In
what follows, we will solve equations of the form \[
\mathbf{P}\mathbf{A}^{(0)}\mathbf{f}=Pg,\]
 where vector components of $\mathbf{f}$ are required to be solenoidal;
this equation is thus equivalent to \begin{equation}
\mathbf{P}\mathbf{A}^{(0)}\mathbf{P}f=\mathbf{P}g.\label{eq:edlin}\end{equation}
 By the Fredholm alternative \citep{Rie90}, a solution of \eqref{eq:edlin}
exists if and only if $\mathbf{P}g$ is orthogonal to the kernel of
$(\mathbf{P}\mathbf{A}^{(0)}\mathbf{P})^{*}$, where $*$ denotes
the adjoint operator. In other words, the solvability condition for
\eqref{eq:edlin} is $\langle\mathbf{P}g,c\rangle=0$, where $\mathbf{c}$
is any vector in $\textrm{ker}(\mathbf{P}\mathbf{A}^{(0)^{*}}\mathbf{P})$
and $\langle\cdot,\cdot\rangle$ denotes the ${\cal L}^{2}$ inner
product. As usual, the adjoint operator $\mathbf{A}^{(0)^{*}}$ can
be derived performing integration by parts in the identity $\langle\mathbf{A}^{(0)^{*}}\mathbf{W}_{1},\mathbf{W}_{2}\rangle=\langle\mathbf{W}_{1},\mathbf{A}^{(0)}\mathbf{W}_{2}\rangle$.
In the present case \[
\mathbf{A}^{(0)^{*}}=\left[\begin{array}{lllll}
\nu\partial^{2}-\partial\,\times(\tilde{\mathbf{V}}\times\bullet)+(\partial\times\tilde{\mathbf{V}})\times &  & \tilde{\mathbf{H}}\times(\partial\times\bullet) &  & -\mathbf{e}_{3}\delta T+\tilde{\theta}\partial\\
\partial\times(\tilde{\mathbf{H}}\times\bullet)-(\partial\times\tilde{\mathbf{H}})\times &  & \eta\partial^{2}-\tilde{\mathbf{V}}\times(\partial\times\bullet) &  & \sigma\partial\times(\partial\times\tilde{\mathbf{H}})-\sigma(\partial\times\tilde{\mathbf{H}})\times(\partial\bullet)\\
-\alpha\mathbf{G}\,\cdot &  & 0 &  & k\partial^{2}+\tilde{\mathbf{V}}\cdot\partial\end{array}\right].\]
 Boundary conditions for the vector fields in the domain of $\mathbf{A}^{(0)^{*}}$
can be found from the condition that boundary surface integrals, emerging
in integration by parts in the scalar product $\langle\mathbf{W}_{1},\mathbf{A}^{(0)}\mathbf{W}_{2}\rangle$,
vanish. It can be verified that the boundary conditions that we assume
hold for vector fields in the domain of the adjoint operator as well.

Our construction relies on the existence of vector fields in $\textrm{ker}(\mathbf{P}\mathbf{A}^{(0)}\mathbf{P})$
with non-vanishing average horizontal components of the flow and magnetic
field (see a detailed discussion in \citep{Zhe05}). It can be easily
established that the dimension of the subspace of $\textrm{ker}(\mathbf{P}\mathbf{A}^{(0)}\mathbf{P})$,
consisting of such vectors, is equal to the dimension of the subspace
of $\textrm{ker}(\mathbf{P}\mathbf{A}^{(0)^{*}}\mathbf{P})$ consisting
of vectors with non-zero horizontal space averages. If $\mathbf{c}=\left[\begin{array}{c}
\mathbf{c}^{V}\\
\mathbf{c}^{H}\\
0\end{array}\right]$, where $\mathbf{c}^{V}$ and $\mathbf{c}^{H}$ are any constant horizontal
vectors, evidently $\mathbf{P}\mathbf{A}^{(0)^{*}}\mathbf{P}c=0$,
since $\mathbf{A}^{(0)^{*}}\mathbf{c}$ is a gradient. Such constant
vector fields satisfy the boundary conditions under consideration;
therefore any such $\mathbf{c}$ belongs to $\textrm{ker}(\mathbf{P}\mathbf{A}^{(0)^{*}}\mathbf{P})$.
Thus, $\textrm{ker}(\mathbf{P}\mathbf{A}^{(0)}\mathbf{P})$ is at
least four-dimensional. In what follows, we assume that the dimension
is four, which is generically the case, and thus any non-zero vector
from $\textrm{ker}(\mathbf{P}\mathbf{A}^{(0)}\mathbf{P})$ has non-zero
horizontal averages of the flow and/or magnetic field components.
Then the solvability condition for \eqref{eq:edlin} consists of orthogonality
of $\mathbf{g}^{V}$ and $\mathbf{g}^{H}$ to constant horizontal
vectors, i.e. the horizontal components of $\langle\mathbf{g}^{V}\rangle$
and $\langle\mathbf{g}^{H}\rangle$ must vanish.

\subsection{Equations at order 0\label{sub:Equations-at-order-0}}

Decoupling of the large- and short-scale behaviour is evident in the
form of solutions of the equations emerging at orders 0 and 1 in the
hierarchy. Since $\mathbf{A}^{(0)}$ is an operator in the fast variables,
short-scale variation of these solutions is exclusively due to their
multiplicative dependence on solutions of \eqref{eq:order0}-\eqref{eq:order0sh}
and \eqref{eq:order1}-\eqref{eq:order1sh} -- the so-called auxiliary
problems, which are partial differential equations in the fast variables.

Performing integration and using the boundary conditions, we find
that the horizontal components of the flow and magnetic field components
of $\langle\mathbf{A}^{(0)}\mathbf{W}^{(0)}\rangle$ vanish. Since
we seek solutions where $\langle\mathbf{V}^{(0)}\rangle$ and $\langle\mathbf{H}^{(0)}\rangle$
do not vanish simultaneously, averaging of \eqref{eq:order0} implies
$\lambda_{0}=0$. The problem to solve becomes \begin{equation}
\mathbf{A}^{(0)}\mathbf{W}^{(0)}=\left[\begin{array}{c}
\partial p^{(0)}\\
\mathbf{0}\\
0\end{array}\right].\label{eq:aux0_raw}\end{equation}

By linearity, \begin{equation}
\mathbf{W}^{(0)}=\sum_{i=1}^{4}a_{i}\,\mathbf{S}_{i},\quad\{ p^{(0)}\}=\sum_{i=1}^{4}a_{i}\, S_{i}^{p}.\label{eq:sol0}\end{equation}
 The fields $\mathbf{S}_{i}=\left[\begin{array}{c}
\mathbf{S}_{i}^{\mathbf{V}}\\
\mathbf{S}_{i}^{\mathbf{H}}\\
S_{i}^{\theta}\end{array}\right]$ and $S_{i}^{p}$ are linearly independent solutions of the problems
\begin{equation}
\begin{array}{l}
\mathbf{A}^{(0)}\mathbf{S}_{i}=\left[\begin{array}{c}
\partial S_{i}^{p}\\
\mathbf{0}\\
0\end{array}\right],\\
\partial\cdot\mathbf{S}_{i}^{\mathbf{V}}=0,\\
\partial\cdot\mathbf{S}_{i}^{\mathbf{H}}=0,\end{array}\label{eq:aux0}\end{equation}
 i.e. $\mathbf{S}_{i}$ are linearly independent fields in $\textrm{ker}(\mathbf{P}\mathbf{A}^{(0)}\mathbf{P})$.
The gradient part of \eqref{eq:aux0} satisfies \begin{equation}
\partial^{2}S_{i}^{p}=\partial\cdot\left(\mathbf{A}^{(0)}\mathbf{S}_{i}\right)^{\mathbf{V}}.\label{eq:auxp0}\end{equation}
 Note that $\langle p^{(0)}\rangle$ cannot be determined at this
order, since derivatives in the fast variables in the r.h.s. of \eqref{eq:aux0_raw}
eliminate any averages in $p^{(0)}$. $\mathbf{S}_{i}$ and $S_{i}^{p}$
depend only on the fast variables. The coefficients of the linear
combinations (\eqref{eq:sol0}), $a_{i}$, depend only on the slow
variables.

The four possible averages (constant mean fields) in $\textrm{ker}(\mathbf{P}\mathbf{A}^{(0)}\mathbf{P})$,
selected by the boundary conditions, are \[
\begin{array}{cc}
\langle\mathbf{S}_{1}\rangle=\left[\begin{array}{c}
\mathbf{e}_{1}\\
\mathbf{0}\\
0\end{array}\right], & \langle\mathbf{S}_{2}\rangle=\left[\begin{array}{c}
\mathbf{e}_{2}\\
\mathbf{0}\\
0\end{array}\right],\end{array}\]
 \[
\begin{array}{cc}
\langle\mathbf{S}_{3}\rangle=\left[\begin{array}{c}
\mathbf{0}\\
\mathbf{e}_{1}\\
0\end{array}\right], & \langle\mathbf{S}_{4}\rangle=\left[\begin{array}{c}
\mathbf{0}\\
\mathbf{e}_{2}\\
0\end{array}\right]\end{array}.\]
 Hence, the four \emph{auxiliary problems at order 0} are \begin{equation}
\mathbf{A}^{(0)}\{\mathbf{S}_{1}\}=\left[\begin{array}{c}
\partial(S_{1}^{p}-\tilde{\mathbf{V}}_{1})+\partial_{1}\tilde{\mathbf{V}}\\
\partial_{1}\tilde{\mathbf{H}}\\
\partial_{1}\tilde{\theta}\end{array}\right],\label{eq:aux01}\end{equation}
 \begin{equation}
\mathbf{A}^{(0)}\{\mathbf{S}_{2}\}=\left[\begin{array}{c}
\partial(S_{2}^{p}-\tilde{\mathbf{V}}_{2})+\partial_{2}\tilde{\mathbf{V}}\\
\partial_{2}\tilde{\mathbf{H}}\\
\partial_{2}\tilde{\theta}\end{array}\right],\label{eq:aux02}\end{equation}
 \begin{equation}
\mathbf{A}^{(0)}\{\mathbf{S}_{3}\}=\left[\begin{array}{c}
\partial(S_{3}^{p}+\tilde{\mathbf{H}}_{1})-\partial_{1}\tilde{\mathbf{H}}\\
-\partial_{1}\tilde{\mathbf{V}}\\
0\end{array}\right],\label{eq:aux03}\end{equation}
 \begin{equation}
\mathbf{A}^{(0)}\{\mathbf{S}_{4}\}=\left[\begin{array}{c}
\partial(S_{4}^{p}+\tilde{\mathbf{H}}_{2})-\partial_{2}\tilde{\mathbf{H}}\\
-\partial_{2}\tilde{\mathbf{V}}\\
0\end{array}\right],\label{eq:aux04}\end{equation}
 with $S_{i}^{p}$ given by \eqref{eq:auxp0}.

Solenoidal parts of the right hand sides of \eqref{eq:aux01}-\eqref{eq:aux04}
are anti-symmetric and, since $\mathbf{A}^{(0)}$ preserves the symmetry
of fields, $\mathbf{S}_{i}$ are anti-symmetric. Evidently, the order
0 auxiliary problems are of the form of \eqref{eq:edlin}, and their
solvability follows immediately from the periodicity of the CHM steady
state $\tilde{\mathbf{V}}$, $\tilde{\mathbf{H}}$, $\tilde{\theta}$
in horizontal directions.

\subsection{Equations at order 1\label{sub:Equations-at-order-1}}

Averaging of \eqref{eq:order1} yields \begin{equation}
\left\langle \mathbf{A}^{(0)}\mathbf{W}^{(1)}\right\rangle =-\left\langle \mathbf{A}^{(1)}\mathbf{W}^{(0)}\right\rangle +\lambda_{1}\langle\mathbf{W}^{(0)}\rangle+\left[\begin{array}{c}
\nabla\langle p^{(0)}\rangle\\
\mathbf{0}\\
0\end{array}\right].\label{eq:av_aux1_raw}\end{equation}
 As in order 0, horizontal parts of the flow and magnetic field components
of $\langle\mathbf{A}^{(0)}\mathbf{W}^{(1)}\rangle$ are zero. The
same holds for $\langle\mathbf{A}^{(1)}\mathbf{W}^{(0)}\rangle$ (this
can be shown, integrating directly the terms of the form of spatial
derivatives and using the boundary conditions, and exploiting the
symmetry of the perturbed CHM steady state and anti-symmetry in the
fast variables of $\mathbf{V}^{(0)}$ and $\mathbf{H}^{(0)}$ when
considering mean horizontal parts of the terms of the form of $\tilde{\mathbf{V}}\times(\nabla\times\mathbf{V}^{(0)})$).
\eqref{eq:av_aux1_raw} becomes \[
\lambda_{1}\langle\mathbf{W}^{(0)}\rangle+\left[\begin{array}{c}
\nabla\langle p^{(0)}\rangle\\
\mathbf{0}\\
0\end{array}\right]=\left[\begin{array}{c}
\mathbf{0}\\
\mathbf{0}\\
0\end{array}\right].\]
 Thus, if $\lambda_{1}\ne0$, then $\langle\mathbf{H}^{(0)}\rangle=\mathbf{0}$
and $\langle\mathbf{V}^{(0)}\rangle=-\nabla\langle p^{(0)}\rangle/\lambda_{1}$.
However, averaging of \eqref{eq:order1sv} over the fast variables
yields $\nabla\cdot\langle\mathbf{V}^{(0)}\rangle=0$, with $\langle\mathbf{V}^{(0)}\rangle$
and $\nabla\langle p^{(0)}\rangle$ belonging thereby to orthogonal
subspaces. This implies $\langle\mathbf{V}^{(0)}\rangle=\mathbf{0}$,
which contradicts the original assumption that $\langle\mathbf{V}^{(0)}\rangle$
and $\langle\mathbf{H}^{(0)}\rangle$ do not vanish simultaneously.
Therefore, $\lambda_{1}=0$, $\langle p^{(0)}\rangle=0$ and \eqref{eq:order1}
reduces to \begin{equation}
\mathbf{A}^{(0)}\mathbf{W}^{(1)}=-\mathbf{A}^{(1)}\mathbf{W}^{(0)}+\left[\begin{array}{c}
\nabla p^{(0)}\\
\mathbf{0}\\
0\end{array}\right]+\left[\begin{array}{c}
\partial p^{(1)}\\
\mathbf{0}\\
0\end{array}\right].\label{eq:aux1_raw}\end{equation}
 From \eqref{eq:sol0} we find \[
-\mathbf{A}^{(1)}\mathbf{W}^{(0)}+\left[\begin{array}{c}
\nabla p^{(0)}\\
\mathbf{0}\\
0\end{array}\right]=\sum_{i=1}^{4}\sum_{j=1}^{2}\mathbf{M}_{ij}\nabla_{j}a_{i},\]
 where \[
\mathbf{M}_{ij}=-\mathbf{B}_{j}\mathbf{S}_{i}+\left[\begin{array}{c}
\mathbf{e}_{j}S_{i}^{p}\\
\mathbf{0}\\
0\end{array}\right]\]
 and\begin{equation}
\mathbf{B}_{j}=\left[\begin{array}{lclcl}
2\nu\partial_{j}+\mathbf{e}_{j}\tilde{\mathbf{V}}\cdot\bullet-\tilde{V}_{j} &  & -\mathbf{e}_{j}\tilde{\mathbf{H}}\cdot\bullet+\tilde{H}_{j} &  & \mathbf{0}\\
-\tilde{\mathbf{H}}\mathbf{e}_{j}\cdot\bullet+\tilde{H}_{j} &  & 2\eta\partial_{j}+\tilde{\mathbf{V}}\mathbf{e}_{j}\cdot\bullet-\tilde{V}_{j} &  & \mathbf{0}\\
0 &  & \sum_{k=1}^{3}\left(\partial_{j}\tilde{H}_{k}-\partial_{k}\tilde{H}_{j}\right)\mathbf{e}_{k}\cdot &  & 2k\partial_{j}-\tilde{V}_{j}\end{array}\right].\label{eq:bj}\end{equation}
 The problem at this order reduces to \begin{eqnarray*}
\mathbf{A}^{(0)}\mathbf{W}^{(1)} & = & \sum_{i=1}^{4}\sum_{j=1}^{2}\mathbf{M}_{ij}\nabla_{j}\alpha_{i}+\left[\begin{array}{c}
\partial p^{(1)}\\
\mathbf{0}\\
0\end{array}\right].\end{eqnarray*}
 Therefore, by linearity, \begin{equation}
\mathbf{W}^{(1)}=\sum_{i=1}^{4}\sum_{j=1}^{2}\nabla_{j}a_{i}\bGamma_{ij}+\sum_{i=1}^{4}b_{i}\mathbf{S}_{i},\label{eq:sol1}\end{equation}
 \begin{equation}
\{ p^{(1)}\}=\sum_{i=1}^{4}\sum_{j=1}^{2}\nabla_{j}a_{i}\Gamma_{ij}^{p}+\sum_{i=1}^{4}b_{i}S_{i}^{p},\label{eq:solp1}\end{equation}
 where $\bGamma_{ij}=\left[\begin{array}{c}
\bGamma_{ij}^{\mathbf{V}}\\
\bGamma_{ij}^{\mathbf{H}}\\
\Gamma_{ij}^{\theta}\end{array}\right]$ and $\Gamma_{ij}^{p}$ are mean-free linearly independent solutions
of the \textit{auxiliary problem at order 1}: \begin{equation}
\begin{array}{l}
\mathbf{A}^{(0)}\bGamma_{ij}=\mathbf{M}_{ij}+\left[\begin{array}{c}
\partial\Gamma_{ij}^{p}\\
\mathbf{0}\\
0\end{array}\right],\\
\partial\cdot\bGamma_{ij}^{\mathbf{V}}=-\left(S_{i}^{\mathbf{V}}\right)_{j},\\
\partial\cdot\bGamma_{ij}^{\mathbf{H}}=-\left(S_{i}^{\mathbf{H}}\right)_{j}.\end{array}\label{eq:aux1}\end{equation}
 Taking the divergence of the velocity component, we obtain a Poisson
equation for $\Gamma_{ij}^{p}$: \begin{equation}
\partial^{2}\Gamma_{ij}^{p}=\partial\cdot\left(\mathbf{A}^{(0)}\bGamma_{ij}-\mathbf{M}_{ij}\right)^{\mathbf{V}}.\label{eq:auxp1}\end{equation}
 The average of $p^{(1)}$ cannot be determined at this order, since
derivatives in the fast variables in the r.h.s. of \eqref{eq:aux1_raw}
eliminate it. In \eqref{eq:sol1} and \eqref{eq:solp1}, $b_{i}$
depend only on the slow variables, and the fields $\bGamma_{ij}$
and $\Gamma_{ij}^{p}$ only on the fast ones. It is convenient to
solve \eqref{eq:aux1} in the subspace of $(3+3+1)$-dimensional vector
fields, where vector components are solenoidal. Consider the substitution
\[
\bGamma_{ij}=\bGammap_{ij}+\left[\begin{array}{c}
\partial\Pi_{ij}^{\mathbf{V}}\\
\partial\Pi_{ij}^{\mathbf{H}}\\
0\end{array}\right].\]
 The conditions \[
\partial^{2}\Pi_{ij}^{\mathbf{V}}=-\left(S_{i}^{\mathbf{V}}\right)_{j},\quad\partial^{2}\Pi_{ij}^{\mathbf{H}}=-\left(S_{i}^{\mathbf{H}}\right)_{j}\]
 imply $\partial\cdot\bGammap_{ij}^{\mathbf{V}}=\partial\cdot\bGammap_{ij}^{\mathbf{H}}=0$.
At order 1 we have thus to solve eight equations: \begin{equation}
\mathbf{P}\mathbf{A}^{(0)}\bGammap_{ij}=\mathbf{P}\left(\mathbf{M}_{ij}-\mathbf{A}^{(0)}\left[\begin{array}{c}
\partial\Pi_{ij}^{\mathbf{V}}\\
\partial\Pi_{ij}^{\mathbf{H}}\\
0\end{array}\right]\right)\label{eq:aux1all}\end{equation}
 for $i=1,\dots,4$ and $j=1,2$. Solvability of \eqref{eq:aux1all}
can be easily verified by symmetry arguments, since $\mathbf{B}_{j}$
changes the symmetry of fields.

\subsection{The mean-field equations for the CHM instability mode\label{sub:The-mean-field-equations}}

At order 2 the solvability condition is not trivially satisfied and
yields equations for the large-scale mean components of the instability
mode. We consider orthogonality of the r.h.s. of \eqref{eq:order2}
to $\textrm{ker}\,(\mathbf{P}\mathbf{A}^{(0)^{*}}\mathbf{P})$, i.e.
\begin{equation}
\left\langle \mathbf{C}_{l},\lambda_{2}\mathbf{W}^{(0)}-\mathbf{A}^{(2)}\mathbf{W}^{(0)}-\mathbf{A}^{(1)}\mathbf{W}^{(1)}+\left[\begin{array}{c}
\nabla p^{(1)}\\
\mathbf{0}\\
0\end{array}\right]\right\rangle =0,\label{eq:aux2av}\end{equation}
 $\forall\,\mathbf{C}_{l}\in\textrm{ker}(\mathbf{P}\mathbf{A}^{(0)^{*}}\mathbf{P})$.
From \eqref{eq:sol1}, \[
\mathbf{A}^{(1)}\mathbf{W}^{(1)}=\sum_{i=1}^{4}\sum_{j,k=1}^{2}\mathbf{B}_{k}\bGamma_{ij}\nabla_{k}\nabla_{j}a_{i}.\]
 Since $\mathbf{C}_{l}$ are constant, \eqref{eq:aux2av} is equivalent
to \begin{eqnarray}
 &  & \lambda_{2}\sum_{i=1}^{4}\langle\mathbf{C}_{l},\mathbf{S}_{i}\rangle a_{i}-\sum_{i=1}^{4}\langle\mathbf{C}_{l},\bXi\mathbf{S}_{i}\rangle\nabla^{2}a_{i}-\sum_{i=1}^{4}\sum_{j,k=1}^{2}\langle\mathbf{C}_{l},\mathbf{B}_{k}\bGamma_{ij}\rangle\nabla_{k}\nabla_{j}a_{i}+\langle\mathbf{C}_{l}^{\mathbf{V}},\nabla\langle p^{(1)}\rangle\rangle=0\label{eq:closed2a}\end{eqnarray}
 From the system of equations \eqref{eq:closed2a}, we find $a_{i}$;
then $\mathbf{W}^{(0)}$ is obtained from \eqref{eq:sol0}. Thus,
we have derived a closed set of equations for the leading terms in
the expansions \eqref{eq:Wse}-\eqref{eq:Lambdase} of eigenmodes
and their growth rates. The leading term in the eigenvalue expansion
is $\lambda_{2}$, i.e. $\lambda=O(\varepsilon^{2})$. This growth
rate determines the characteristic slow time scale of the large-scale
dynamics: $T=\varepsilon^{2}t$.

Since only horizontal components of $\mathbf{C}_{l}^{\mathbf{V}}$
and $\mathbf{C}_{l}^{\mathbf{H}}$ can be nonzero for constant vectors
$\mathbf{C}_{l}\in\textrm{ker}(\mathbf{P}\mathbf{A}^{(0)^{*}}\mathbf{P})$
(see section \ref{sub:Solvability-Conditions}), we can choose $\mathbf{C}_{l}=\langle\mathbf{S}_{l}\rangle$.
This implies $\langle\mathbf{S}_{i},\mathbf{C}_{l}\rangle=\delta_{li}$
(here $\delta_{li}$ is the Kronecker symbol). Then \eqref{eq:closed2a}
takes the form \begin{eqnarray}
 &  & \lambda_{2}\sum_{i=1}^{4}\delta_{li}a_{i}+\sum_{i=1}^{4}\sum_{j,k=1}^{2}\langle\mathbf{C}_{l},-\bXi\mathbf{S}_{i}\delta_{jk}-\mathbf{B}_{k}\bGamma_{ij}\rangle\nabla_{k}\nabla_{j}a_{i}\nonumber \\
 &  & +\langle\mathbf{C}_{l}^{\mathbf{V}},\nabla\langle p^{(1)}\rangle\rangle=0.\label{eq:closed2b}\end{eqnarray}
 This is an eigenvalue problem for the second order partial differential
operator with constant coefficients, which is called \textit{combined
eddy diffusivity operator}. It admits Fourier harmonics as eigenfunctions:
\begin{equation}
a_{n}(\mathbf{X})=\hat{a}_{n}(\mathbf{q})e^{i\mathbf{q}\cdot\mathbf{X}},\quad\langle p^{(1)}\rangle=\hat{p}(\mathbf{q})e^{i\mathbf{q}\cdot\mathbf{X}},\label{eq:Fouha}\end{equation}
 where ${\mathbf{q}}=(q_{1},q_{2})$ is an arbitrary unit wavevector
and $n=1,...,4$. Upon substitution we find that the coefficients
$\hat{a}_{n}$ satisfy \begin{equation}
\left[\begin{array}{c}
(\lambda_{2}+\nu)\hat{a}_{1}\\
(\lambda_{2}+\nu)\hat{a}_{2}\\
(\lambda_{2}+\eta)\hat{a}_{3}\\
(\lambda_{2}+\eta)\hat{a}_{4}\end{array}\right]+\mathbf{E}\left[\begin{array}{c}
\hat{a}_{1}\\
\hat{a}_{2}\\
\hat{a}_{3}\\
\hat{a}_{4}\end{array}\right]=-i\hat{p}(\mathbf{q})\left[\begin{array}{c}
q_{1}\\
q_{2}\\
0\\
0\end{array}\right],\label{eq:eveq4}\end{equation}
 where $\mathbf{E}$ is the $4\times4$ matrix \begin{equation}
\mathbf{E}_{li}=\sum_{j,k=1}^{2}q_{k}q_{j}\langle\mathbf{C}_{l},\mathbf{B}_{k}\bGamma_{ij}\rangle.\label{eq:matrix}\end{equation}

Averaging of \eqref{eq:order1sv} and \eqref{eq:order1sh} yields
$\nabla\cdot\left\langle \mathbf{V}^{(0)}\right\rangle =0$ and $\nabla\cdot\left\langle \mathbf{H}^{(0)}\right\rangle =0$.
By virtue of these solenoidality conditions and \eqref{eq:Fouha},
\begin{eqnarray*}
(\hat{a}_{1},\hat{a}_{2}) & = & \hat{a}'_{1}(q_{2},-q_{1}),\\
(\hat{a}_{3},\hat{a}_{4}) & = & \hat{a}'_{2}(q_{2},-q_{1}).\end{eqnarray*}
 Substituting these expressions into \eqref{eq:eveq4} and scalar
multiplying it by $(q_{2},-q_{1};q_{2},-q_{1})$, we reduce \eqref{eq:eveq4}
to an equivalent $2\times2$ eigenvalue problem: \begin{equation}
\left[\begin{array}{c}
(\lambda_{2}+\nu)\hat{a}'_{1}\\
(\lambda_{2}+\eta)\hat{a}'_{2}\end{array}\right]+\mathbf{E}'\left[\begin{array}{c}
\hat{a}'_{1}\\
\hat{a}'_{2}\end{array}\right]=0,\label{eq:eveq2}\end{equation}
 where \[
\mathbf{E}'_{11}=\mathbf{E}_{11}q_{2}^{2}-(\mathbf{E}_{12}+\mathbf{E}_{21})q_{1}q_{2}+\mathbf{E}_{22}q_{1}^{2},\]
 \[
\mathbf{E}'_{12}=\mathbf{E}_{13}q_{2}^{2}-(\mathbf{E}_{14}+\mathbf{E}_{23})q_{1}q_{2}+\mathbf{E}_{24}q_{1}^{2},\]
 \[
\mathbf{E}'_{21}=\mathbf{E}_{31}q_{2}^{2}-(\mathbf{E}_{32}+\mathbf{E}_{41})q_{1}q_{2}+\mathbf{E}_{42}q_{1}^{2},\]
 \[
\mathbf{E}'_{22}=\mathbf{E}_{33}q_{2}^{2}-(\mathbf{E}_{34}+\mathbf{E}_{43})q_{1}q_{2}+\mathbf{E}_{44}q_{1}^{2}.\]

Noting that $\mathbf{q}=(\cos\theta,\sin\theta),$ $\theta\in[0,2\pi],$
we obtain\begin{eqnarray*}
\lambda_{2}^{\pm}(\theta) & = & -\frac{b}{2}\left(1\pm\sqrt{1-\frac{4c}{b}}\right),\end{eqnarray*}
 with $b=\nu+\eta+E'_{11}+E'_{22}$ and $c=\nu\eta+\nu E'_{22}+\eta E'_{11}+E'_{11}E'_{22}-E'_{12}E'_{21}.$
The maximum and minimum growth rates, \begin{equation}
\lambda_{2}^{max}=\max_{\theta\in[0,2\pi]}\max\left\{ \lambda_{2}^{+}(\theta),\lambda_{2}^{-}(\theta)\right\} ,\label{eq:lambda2_max}\end{equation}

\begin{equation}
\lambda_{2}^{min}=\min_{\theta\in[0,2\pi]}\min\left\{ \lambda_{2}^{+}(\theta),\lambda_{2}^{-}(\theta)\right\} ,\label{eq:lambda2_min}\end{equation}
are admitted for $\theta$'s denoted by $\theta^{max}$ and $\theta^{min}$,
respectively.

\section{Numerical Results\label{sec:Numerical-Results}}

The auxiliary problems were solved numerically using pseudo-spectral
methods to evaluate the action of the operators $\mathbf{A}^{(0)}$
\eqref{eq:a0} and $\mathbf{B}_{j}$ \eqref{eq:bj} on the fields.
In the finite direction of the layer, the usual plane wave basis was
replaced by a half period sine or cosine basis, agreeing with the
boundary conditions:\[
f(x,y,z)=\sum_{n_{k_{x}},\, n_{k_{y}},\, n_{k_{z}}}\hat{f}(k_{x},k_{y},k_{z})e^{i(k_{x}x+k_{y}y)}\sin(k_{z}z),\]
 for a scalar function satisfying Dirichlet-kind boundary conditions,
and\[
f(x,y,z)=\sum_{n_{k_{x}},\, n_{k_{y}},\, n_{k_{z}}}\hat{f}(k_{x},k_{y},k_{z})e^{i(k_{x}x+k_{y}y)}\cos(k_{z}z),\]
 for a scalar function satisfying Neumann-kind boundary conditions,
with $k_{x}=2\pi n_{k_{x}}/L_{1},$ $k_{y}=2\pi n_{k_{y}}/L_{2},$
$k_{z}=\pi n_{k_{z}}/L_{3}$ and $n_{k_{x}},$ $n_{k_{y}},$ $n_{k_{z}}\in\mathbb{Z}.$
For each auxiliary problem, a linear system of equations in the Fourier
space was obtained and solved numerically by the conjugate gradients
method \citep{Axe}.

Asymptotic expansions for large molecular diffusivities, as well as
comparison with previous calculations for plan form velocity fields
\citep{Zhe05}, were used to validate the code. As previously stated,
the basic steady state must be stable to short-scale perturbations,
i.e. the dominant eigenvalue ($\lambda_{short}$) of the operator
$\mathbf{A}$ must have a negative real part. The dominant eigenvalue
can be evaluated using the method used in \citep{zhe93nskdbfs} for
perturbations in each of the two symmetry subspaces.

We want to model magnetic instabilities in turbulent convective flows.
For the reasons exposed in the introduction, simulations of fully
turbulent regimes are very resource expensive. Within the scope of
our approach, steady states can be randomly generated with decaying
energy spectrum. Such states satisfy the basic equations for the appropriate
source terms. Usually only a finite number of Fourier harmonics $(k_{min}\le k<k_{max})$
is generated, the remaining being set to $0.$ Applying the appropriate
linear transformations, we make sure that the generated CHM steady
states are solenoidal and possess the required symmetry. The coefficients
are then normalised to obtain the desired energy spectrum and the
norm (r.m.s.) of each field is set to $1$. Algebraic ($E(k)\sim k^{-\xi}$)
or exponential ($E(k)\sim\exp(-\xi k)$) spectra were used in \citep{Zhe01,Zhe03,Zhe05}.

\begin{figure}[H]
\begin{centering}\includegraphics[scale=0.4]{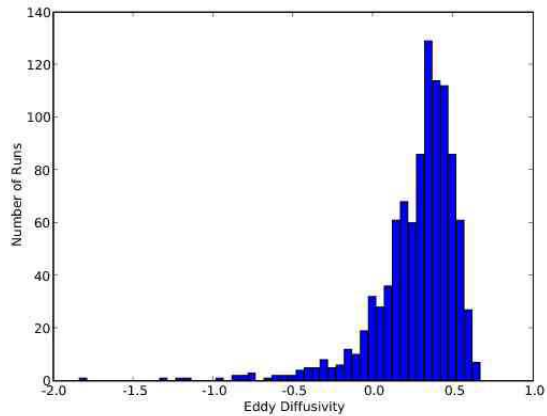}\par\end{centering}

\caption{Statistics of eddy diffusivity\label{fig:Statistics-of-eddy}.}
\end{figure}

Zheligovsky et. al. \citep{Zhe01} found that flows with exponentially
decaying spectra are statistically better dynamos. However, in fully
developed turbulence, the energy spectrum in the inertial range is
known to be algebraic. In this work, all fields have been normalised
to have decaying algebraic energy spectra, with $\xi=4,$ for the
Fourier modes with $0\le k<7.$ Simulations have been carried out
for the periodicity box of size $2\pi\times2\pi\times\pi,$ with the
resolution of $32\times32\times16$ Fourier harmonics. An ensemble
of $1000$ instances of CHM states has been generated. It turns out
that $110$ out of $1000$ generated flows exhibit negative combined
eddy diffusivity (see Fig. \ref{fig:Statistics-of-eddy}). The values
$\nu=\mu=\kappa=0.5$ were chosen that large so that to make sure
that the randomly generated CHM states were stable to short-scale
perturbations. We have directly checked, by computation of the decay
rates of the dominant short-scale modes, that $30$ of the generated
CHM states from our ensemble are indeed stable; for $3$ of them eddy
diffusivity is negative. No instances of CHM states unstable to short-scale
perturbations were found for these values of diffusivities. 

\begin{figure}[H]
\begin{centering}\begin{tabular}{cc}
\includegraphics[width=6.5cm,keepaspectratio]{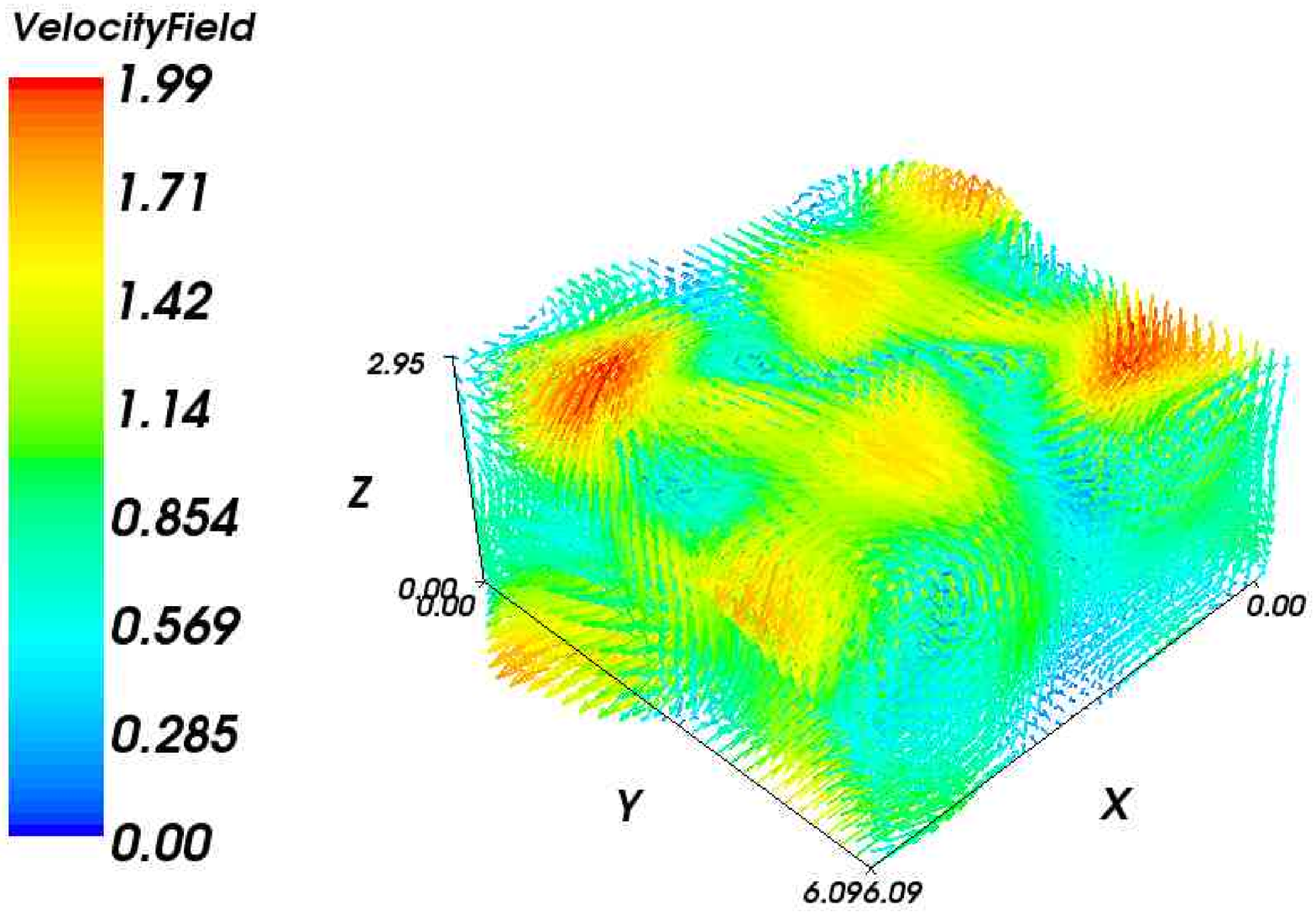}&
\includegraphics[width=6.5cm,keepaspectratio]{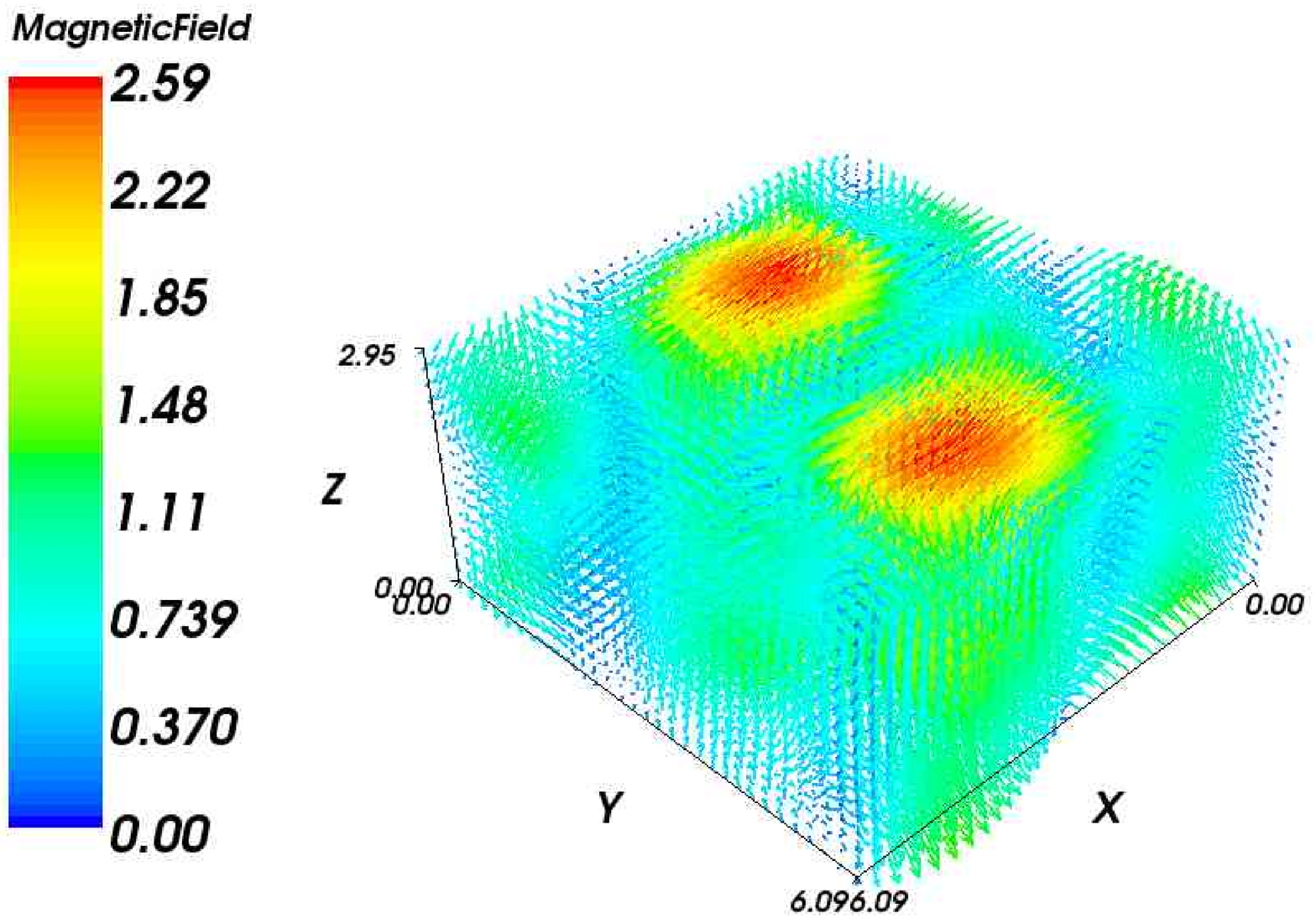}\tabularnewline
Steady state velocity&
Steady state magnetic field\tabularnewline
\includegraphics[width=6.5cm]{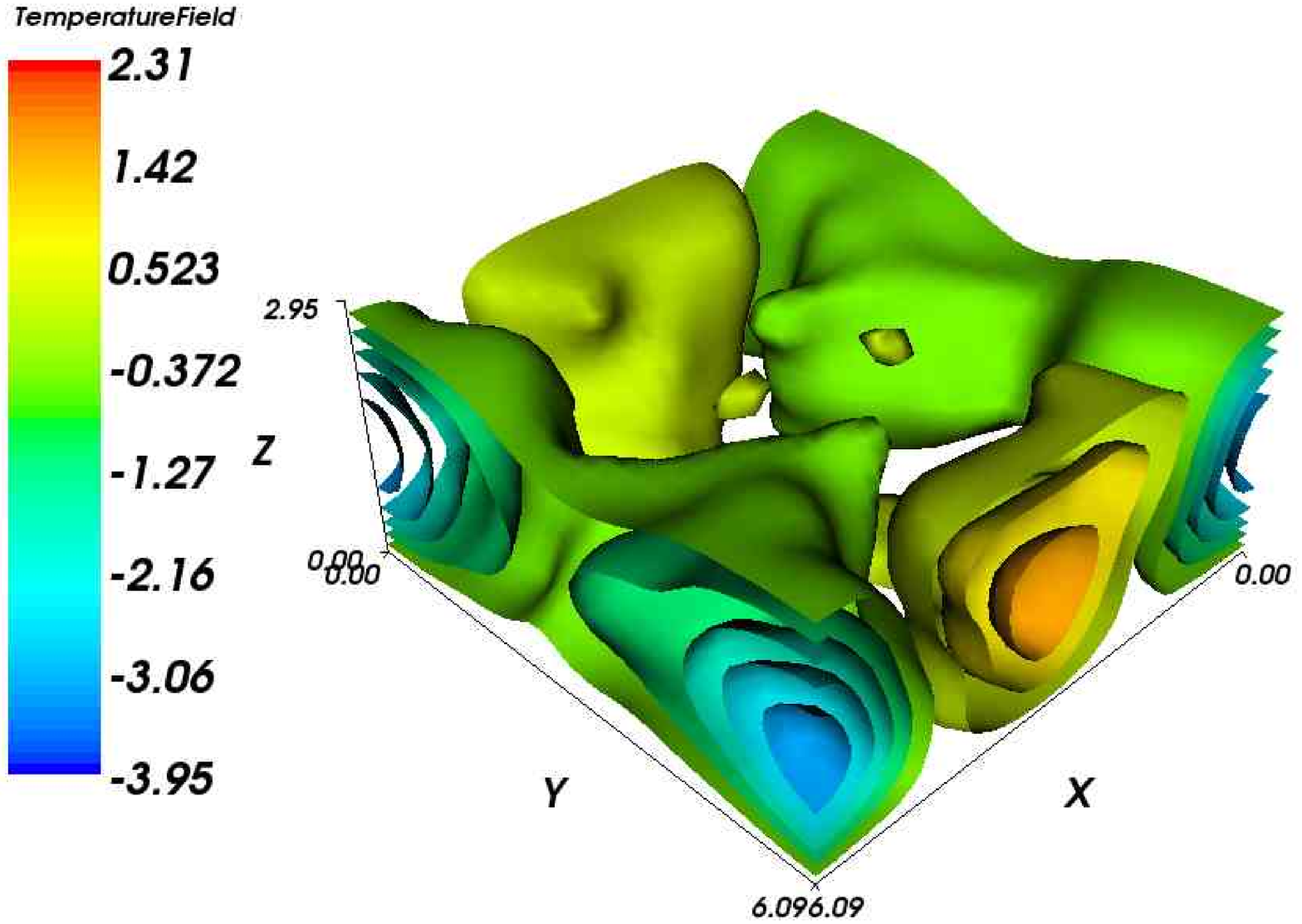}&
\includegraphics[width=6.5cm,keepaspectratio]{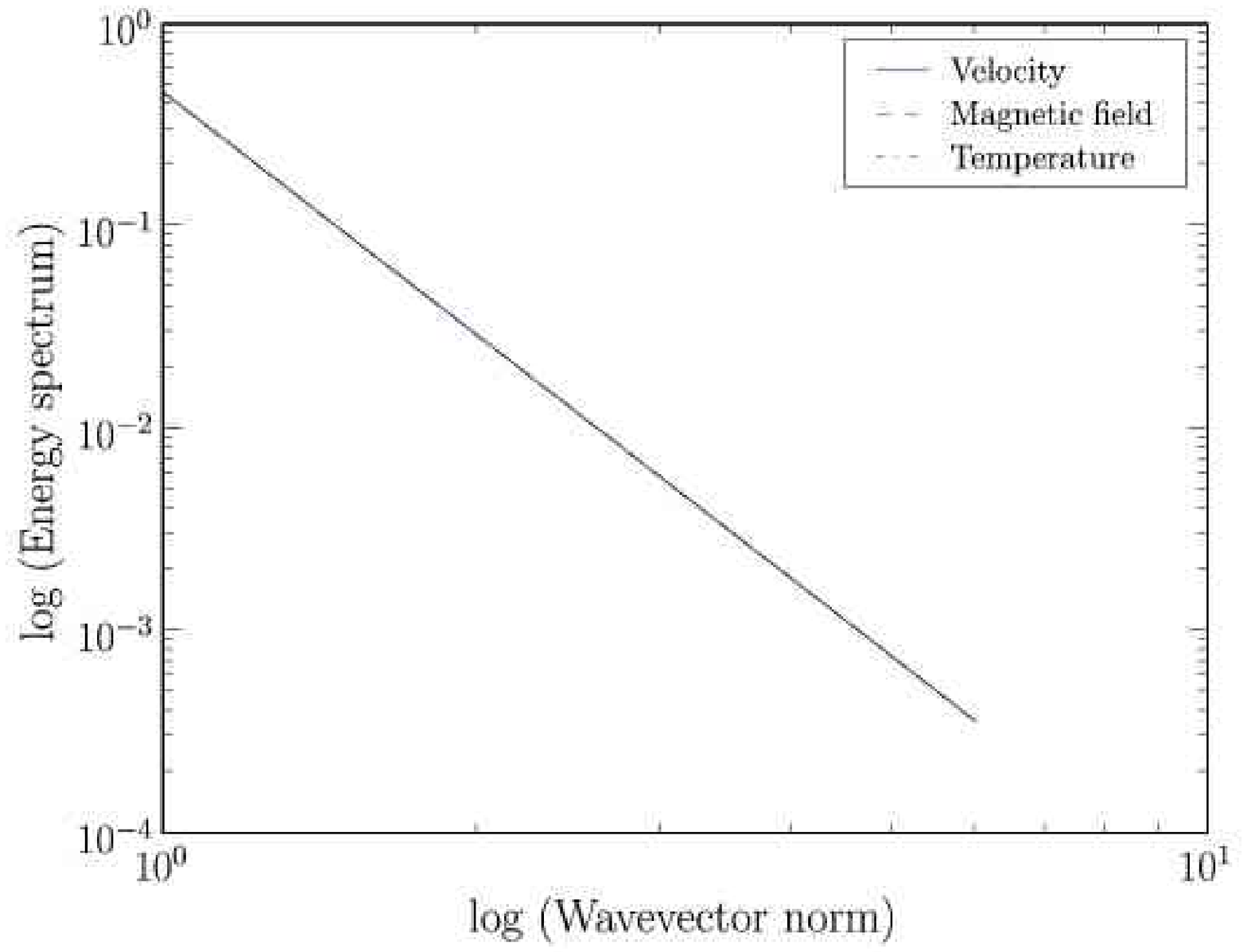}\tabularnewline
Steady state temperature&
Steady state energy spectra\tabularnewline
\end{tabular}\par\end{centering}

\caption{Basic (steady state) fields and corresponding energy spectra. Symmetry
about the $z\mbox{-axis}$ is observed in all of the displayed fields.\label{fig:Steady-state-fields} }
\end{figure}

In figure \ref{fig:Steady-state-fields}, the fields corresponding
to one of the generated steady states are presented. The short-scale
growth rates are $\lambda_{short}=-0.5662$, for symmetric perturbations,
and $\lambda_{short}=-0.05175$, for antisymmetric perturbations.
The maximum and minimum growth rates of large-scale perturbations
are $\lambda_{2}^{max}=1.426$, for $\theta^{max}=3.392$, and $\lambda_{2}^{min}=-1.165$,
for $\theta^{min}=3.593$, respectively. The maximum growth rate is
positive (which corresponds to a negative eigenvalue of the eddy diffusivity
tensor), i.e. a large-scale instability is present. All the auxiliary
problems show a decaying energy spectrum (see figures \ref{fig:S1-fields-and}-\ref{fig:Gamma11-fields-and}
below) and the expected symmetries can be observed.

Spatial short-scale structure of the large-scale eigenmode is defined
by the leading term $\mathbf{W}^{(0)}$ in the expansion \eqref{eq:Wse},
and therefore by the fields $\mathbf{S}_{i}$ (see \eqref{eq:sol0}).
Magnetic field in all of them (see figures \ref{fig:S1-fields-and}-\ref{fig:S4-fields-and})
has the form of distorted \char`\"{}retrograde columns\char`\"{} \citep{spie93}
and concentrates near the horizontal boundaries of the layer. This
kind of structure is reproduced in their linear combination $\mathbf{W}^{(0)}$
for the most unstable mode (see figure \ref{fig:W0-fields-and}).
Such behaviour was originally noticed in the non-linear evolutionary
simulations \citep{spie93} of magnetic field generation by rotating
thermal convection, and it can be attributed to the boundary conditions,
namely, ideal electric conductivity of the boundaries. It is interesting
that this feature is quite robust -- from the formal point of view
our problem is evidently quite different from the one considered in
\citep{spie93}.

\newpage{}

\begin{figure}[H]
\begin{centering}\begin{tabular}{cc}
\includegraphics[width=6.5cm,keepaspectratio]{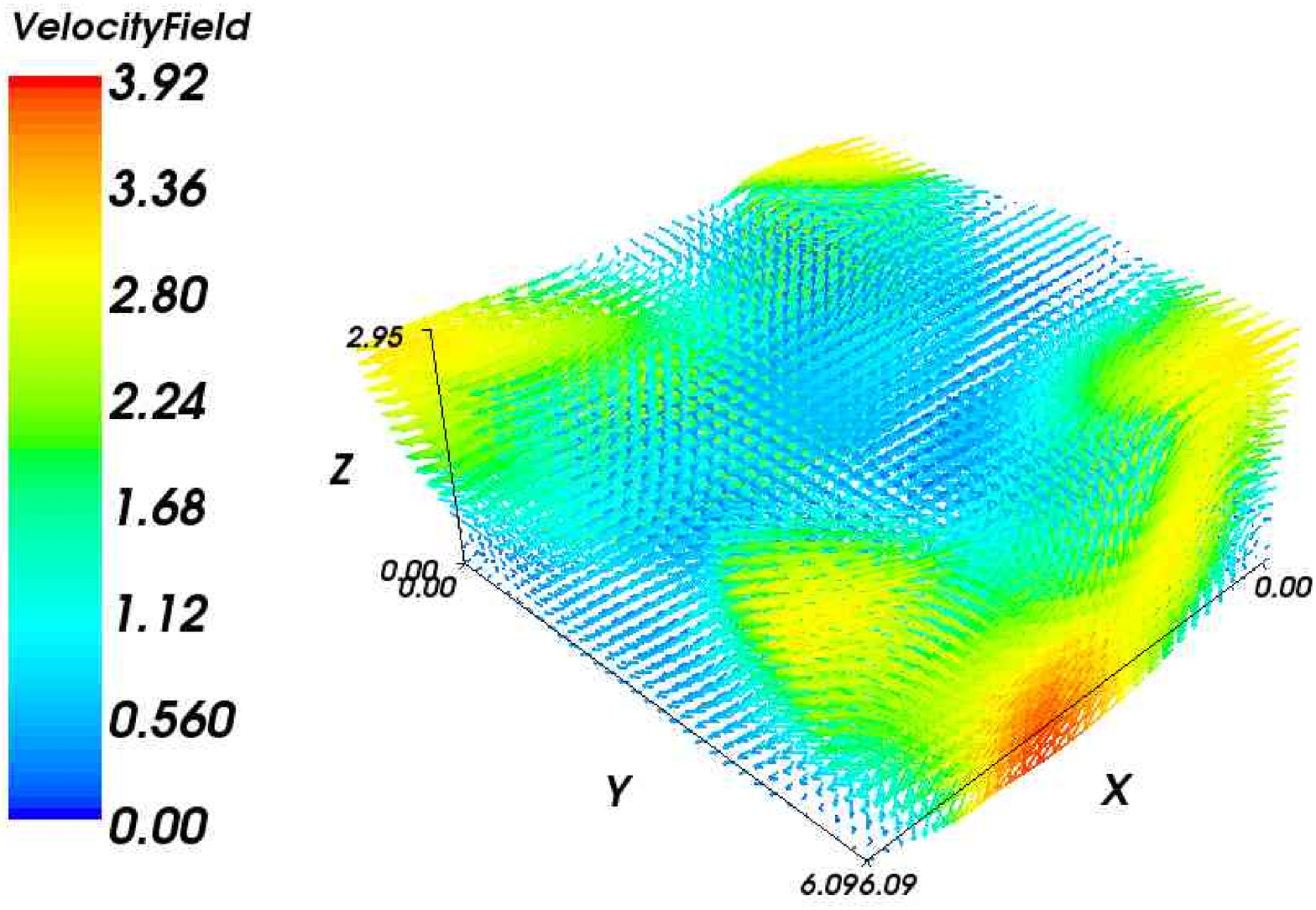}&
\includegraphics[width=6.5cm,keepaspectratio]{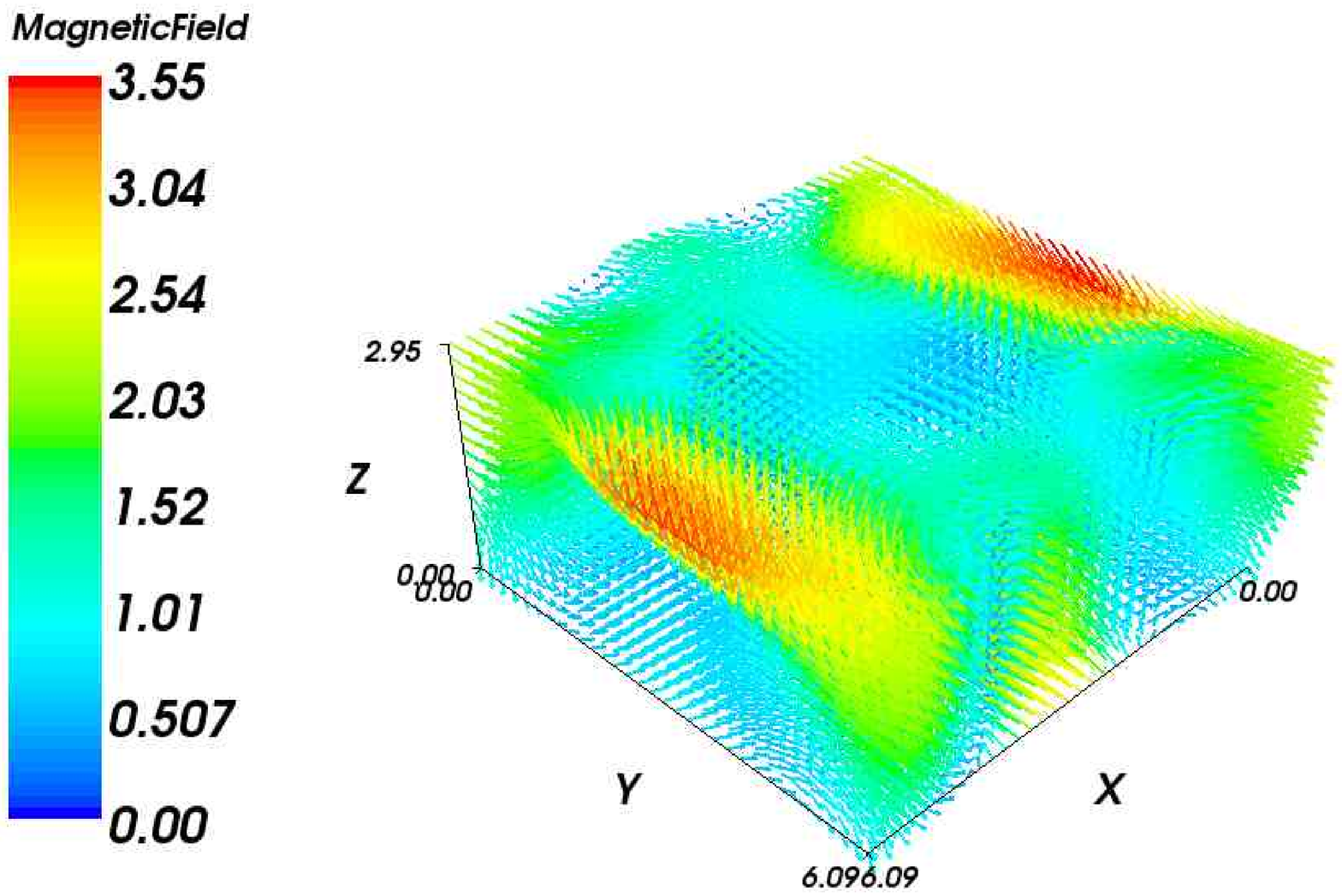}\tabularnewline
$\mathbf{S}_{1}:$ velocity&
$\mathbf{S}_{1}:$ magnetic field\tabularnewline
\includegraphics[width=6.5cm,keepaspectratio]{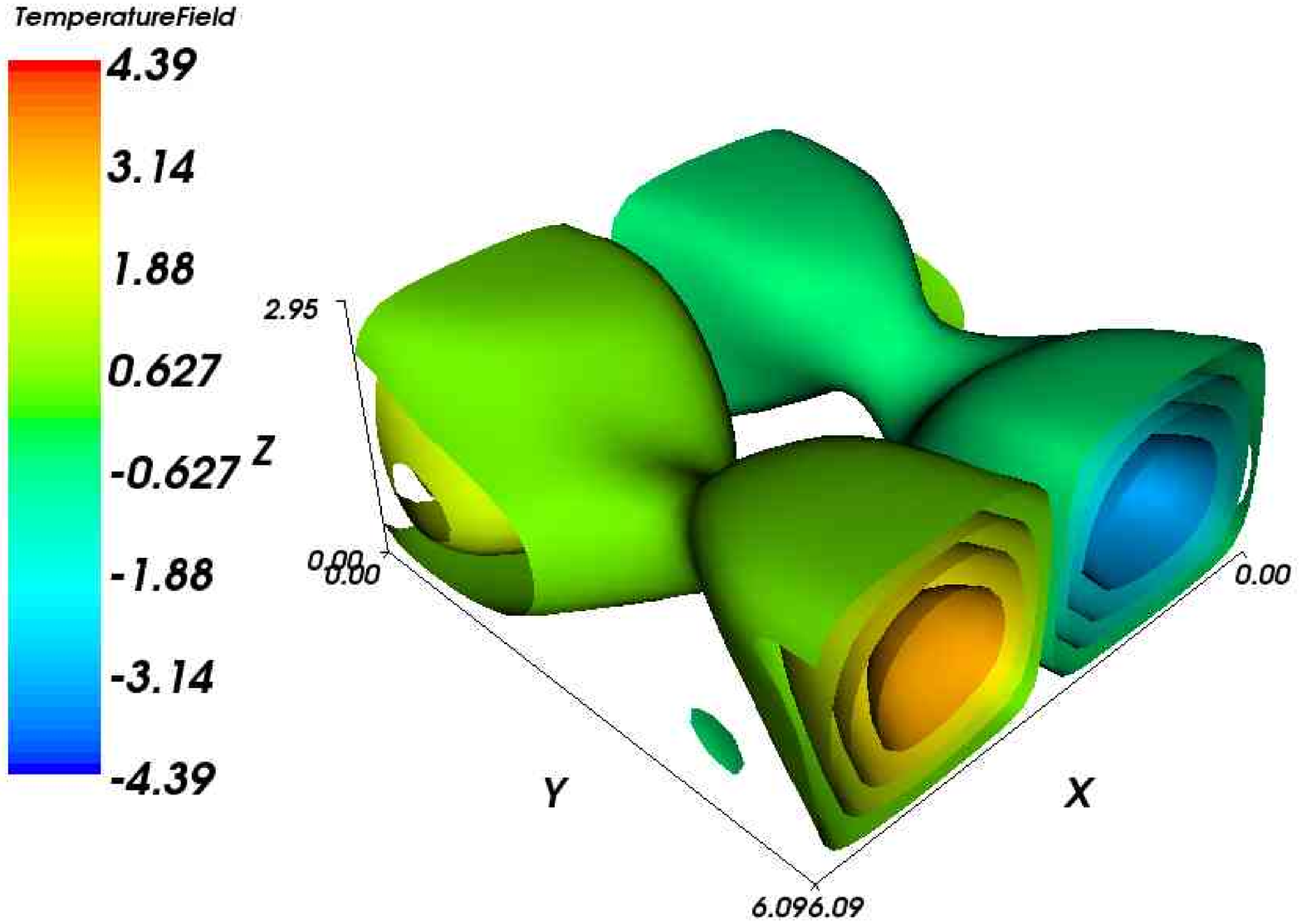}&
\includegraphics[width=6.5cm,keepaspectratio]{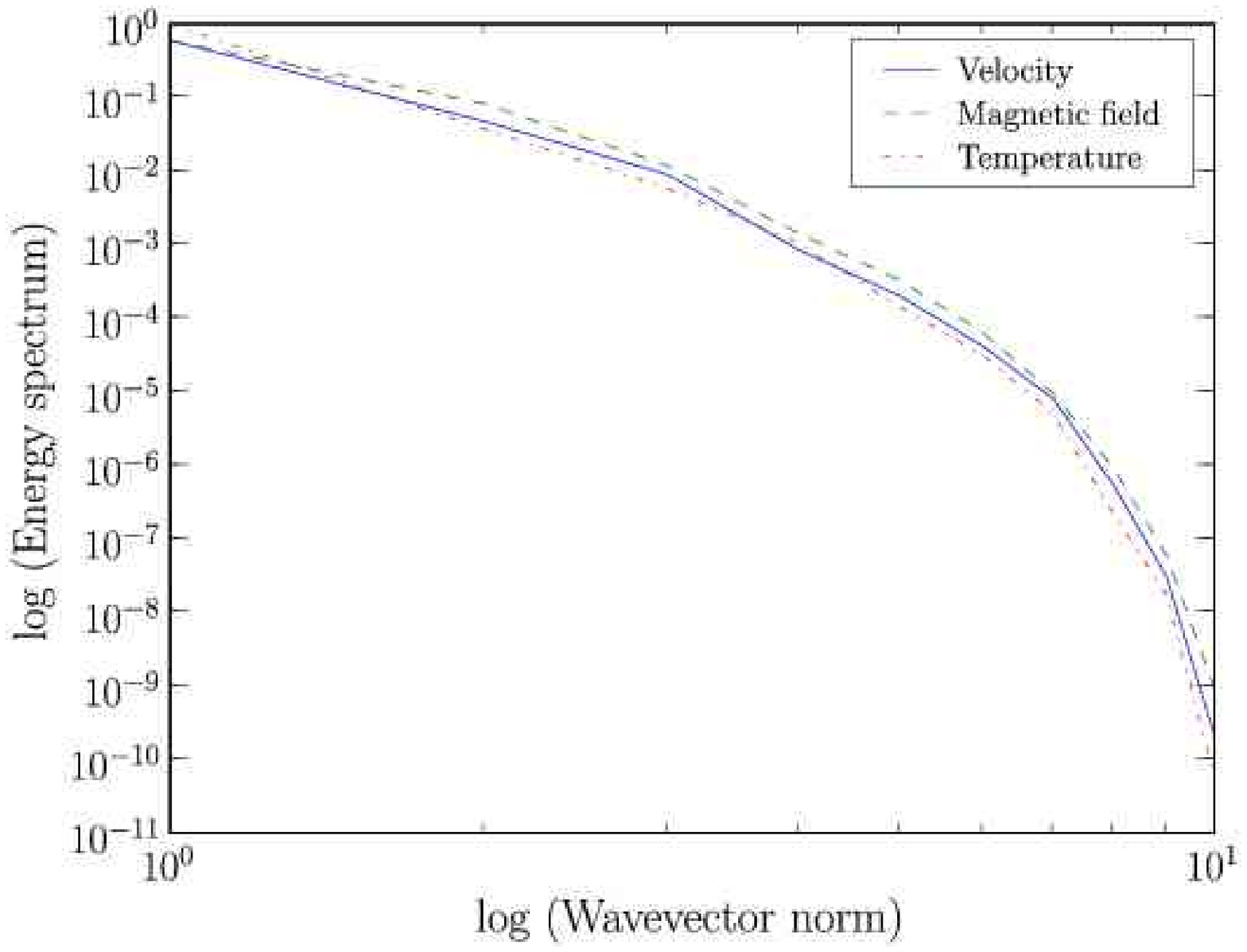}\tabularnewline
$\mathbf{S}_{1}:$ temperature&
$\mathbf{S}_{1}:$ energy spectra\tabularnewline
\end{tabular}\par\end{centering}

\caption{$\mathbf{S}_{1}$ fields and corresponding energy spectra. Anti-symmetry
about the $z\mbox{-axis}$ is observed in all of the displayed fields.\label{fig:S1-fields-and}}
\end{figure}

\begin{figure}[H]
\begin{centering}\begin{tabular}{cc}
\includegraphics[width=6.5cm,keepaspectratio]{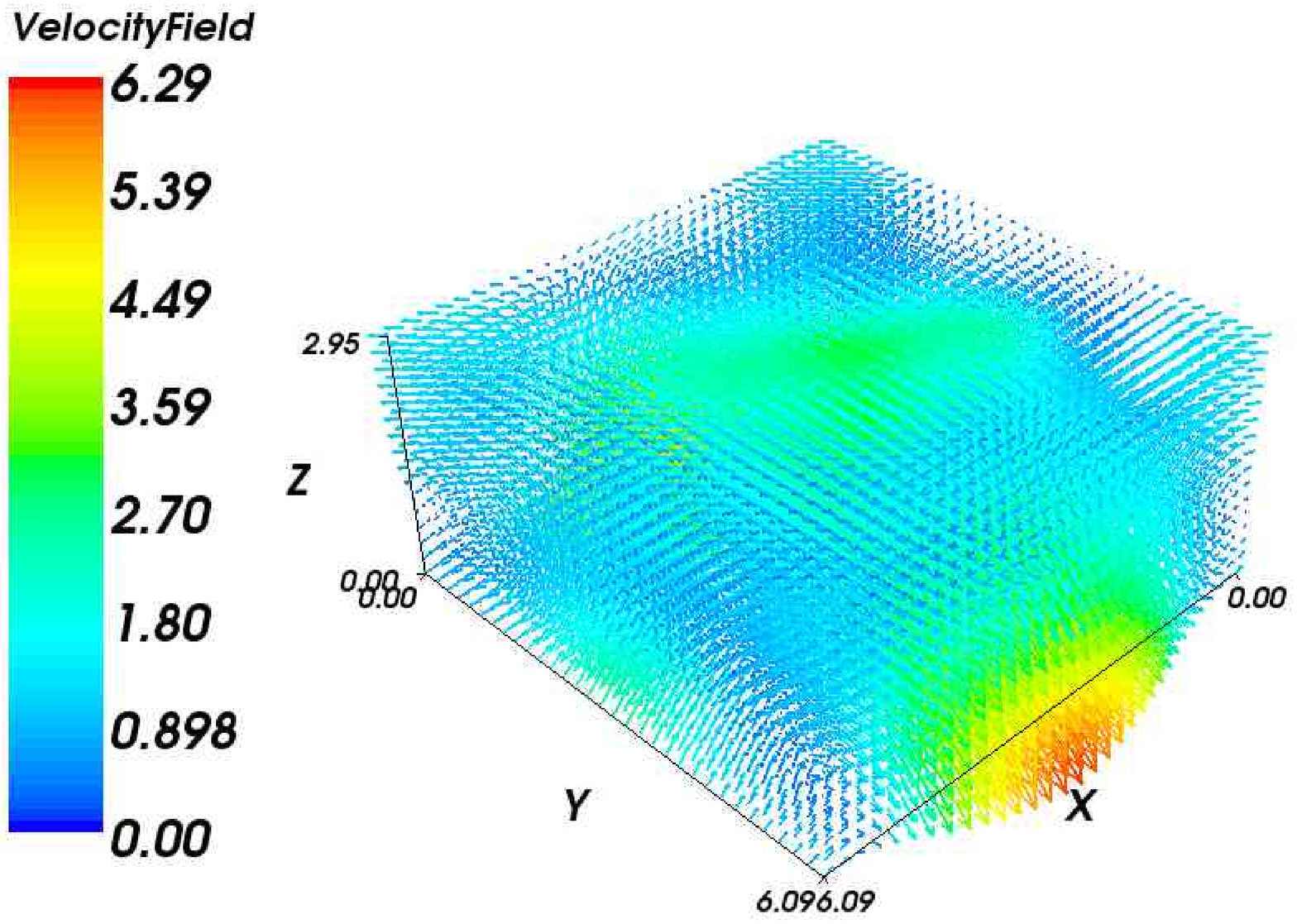}&
\includegraphics[width=6.5cm,keepaspectratio]{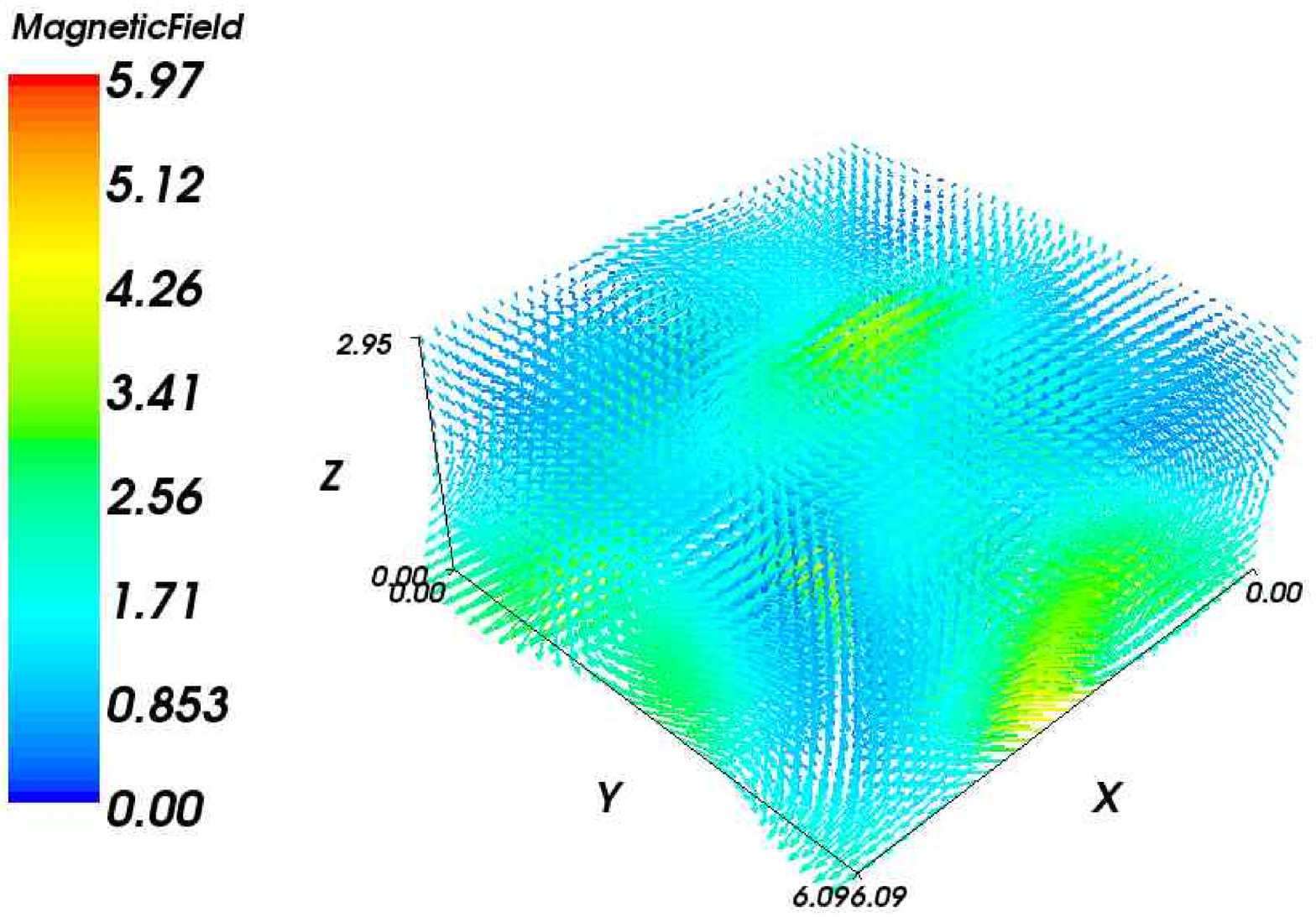}\tabularnewline
$\mathbf{S}_{2}:$ velocity&
$\mathbf{S}_{2}:$ magnetic field\tabularnewline
\includegraphics[width=6.5cm,keepaspectratio]{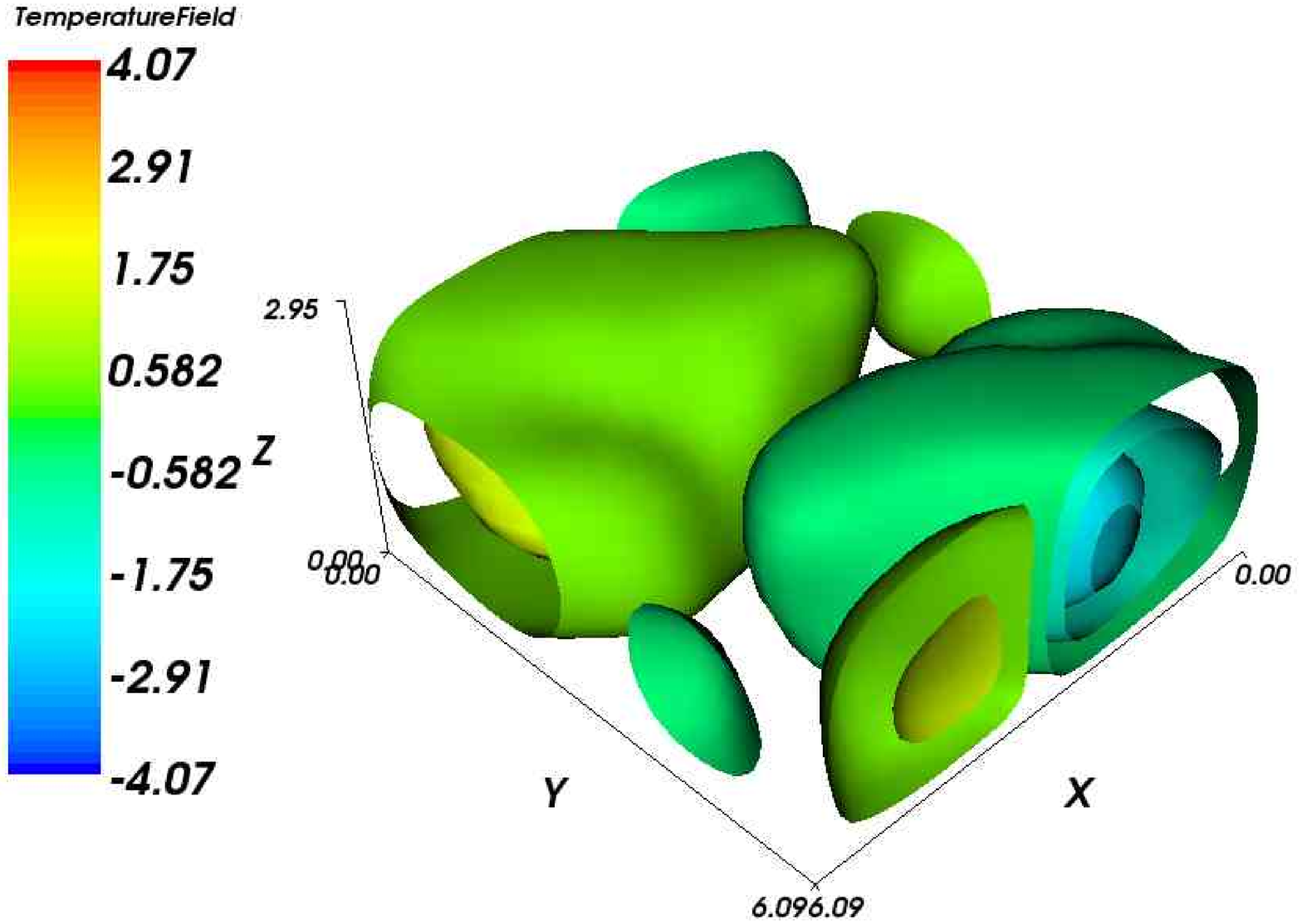}&
\includegraphics[width=6.5cm,keepaspectratio]{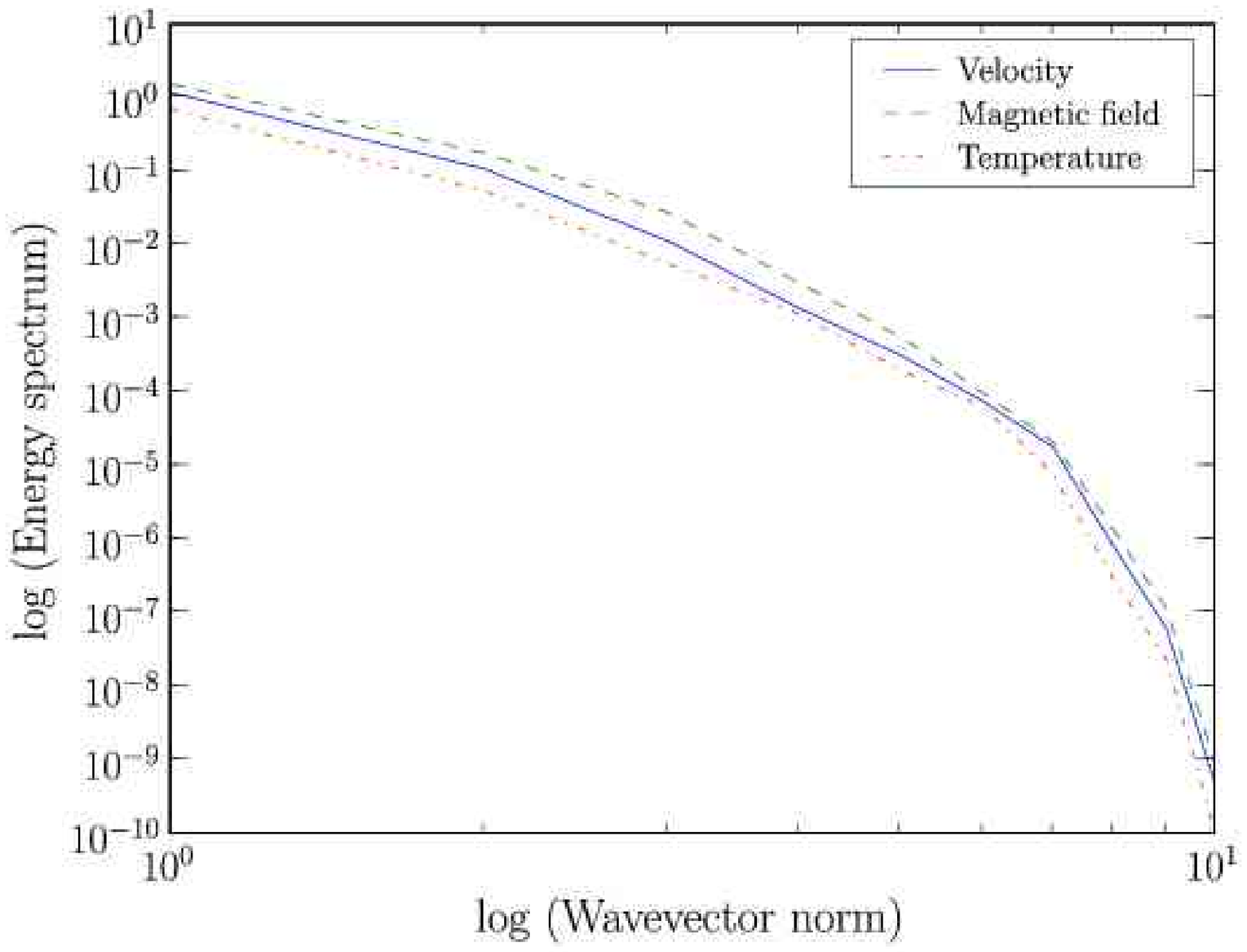}\tabularnewline
$\mathbf{S}_{2}:$ temperature&
$\mathbf{S}_{2}:$ energy spectra\tabularnewline
\end{tabular}\par\end{centering}

\caption{$\mathbf{S}_{2}$ fields and corresponding energy spectra. Anti-symmetry
about the $z\mbox{-axis}$ is observed in all of the displayed fields.\label{fig:S2-fields-and}}
\end{figure}

\begin{figure}[H]
\begin{centering}\begin{tabular}{cc}
\includegraphics[width=6.5cm,keepaspectratio]{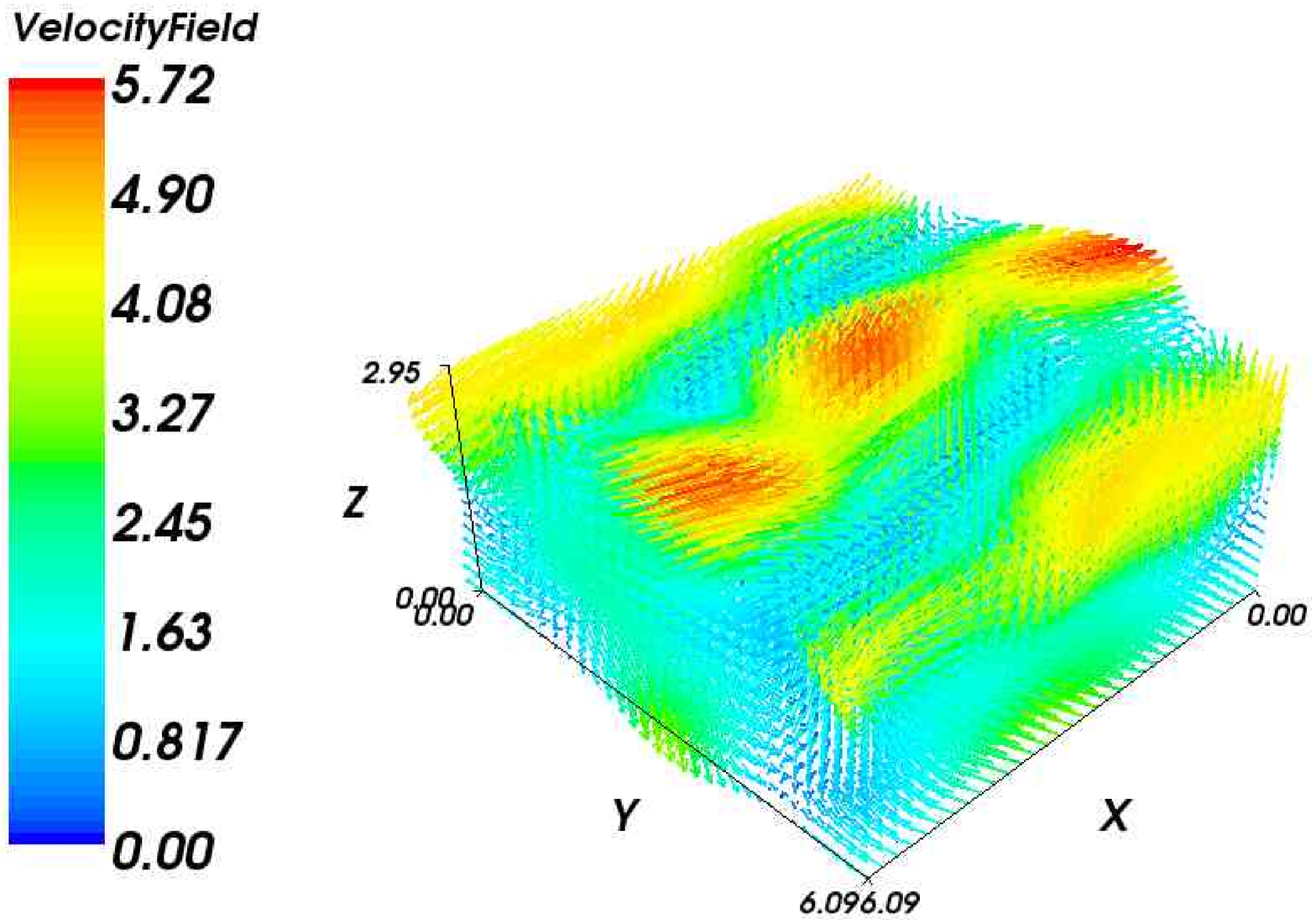}&
\includegraphics[width=6.5cm,keepaspectratio]{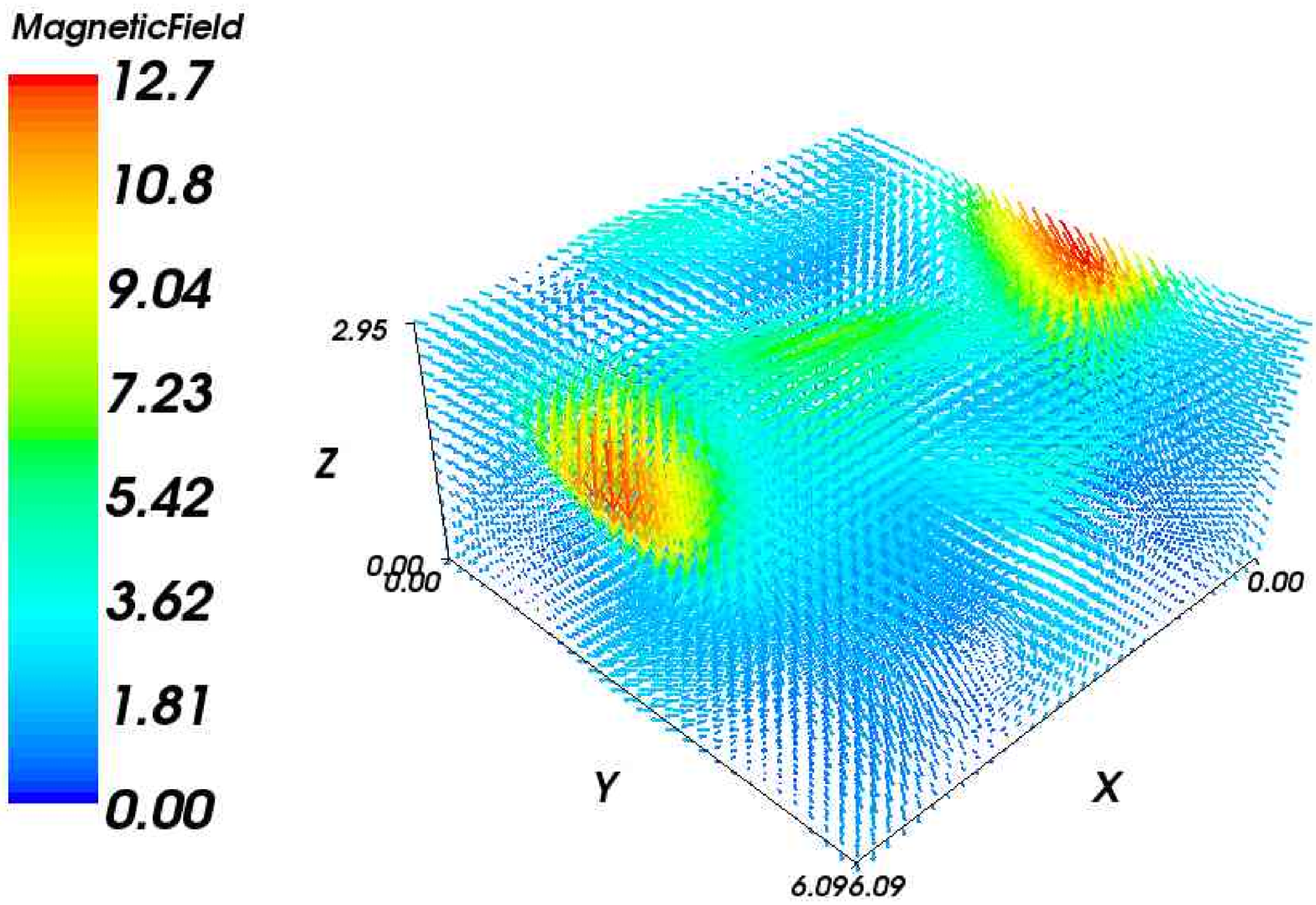}\tabularnewline
$\mathbf{S}_{3}:$ velocity&
$\mathbf{S}_{3}:$ magnetic field\tabularnewline
\includegraphics[width=6.5cm,keepaspectratio]{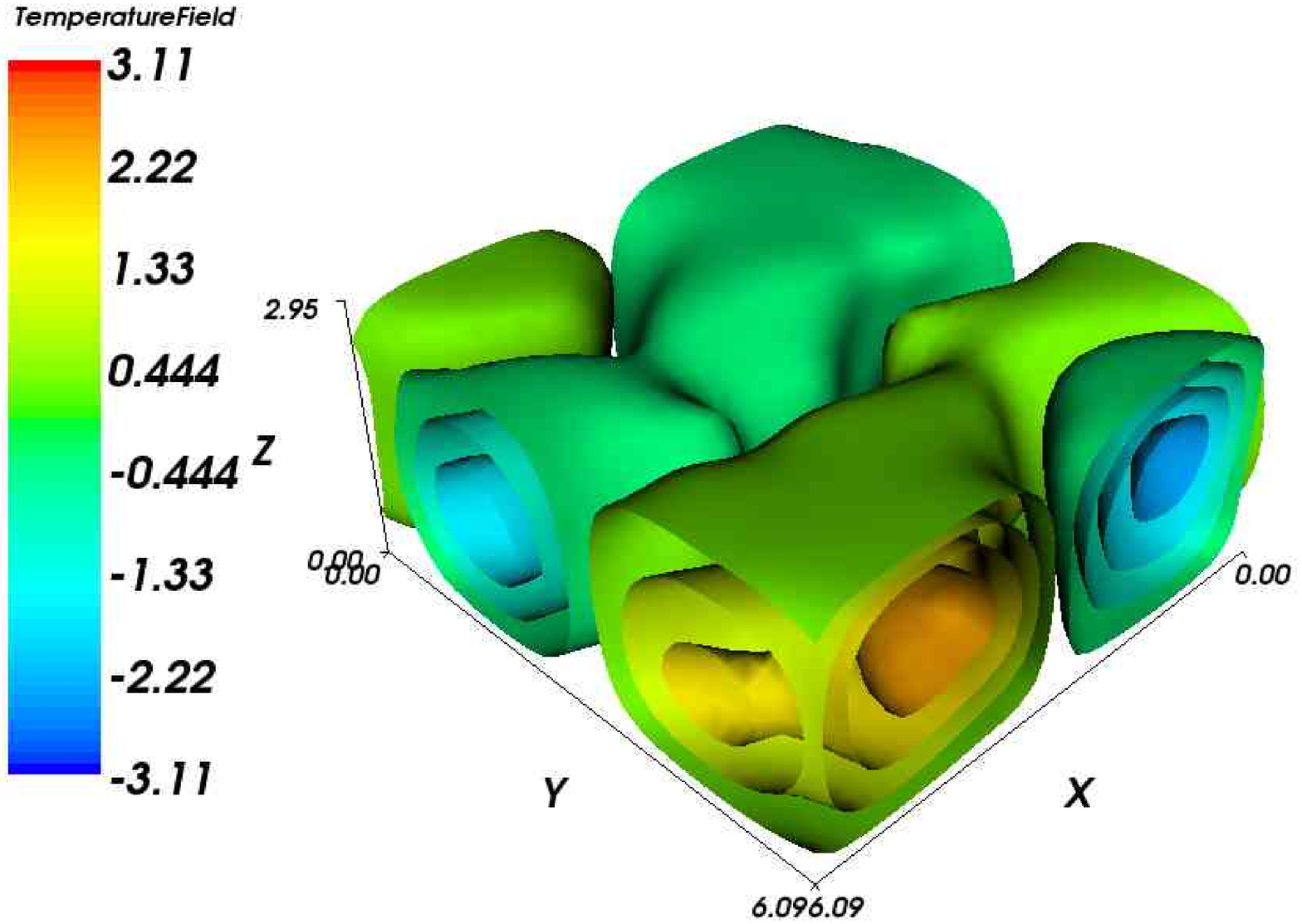}&
\includegraphics[width=6.5cm,keepaspectratio]{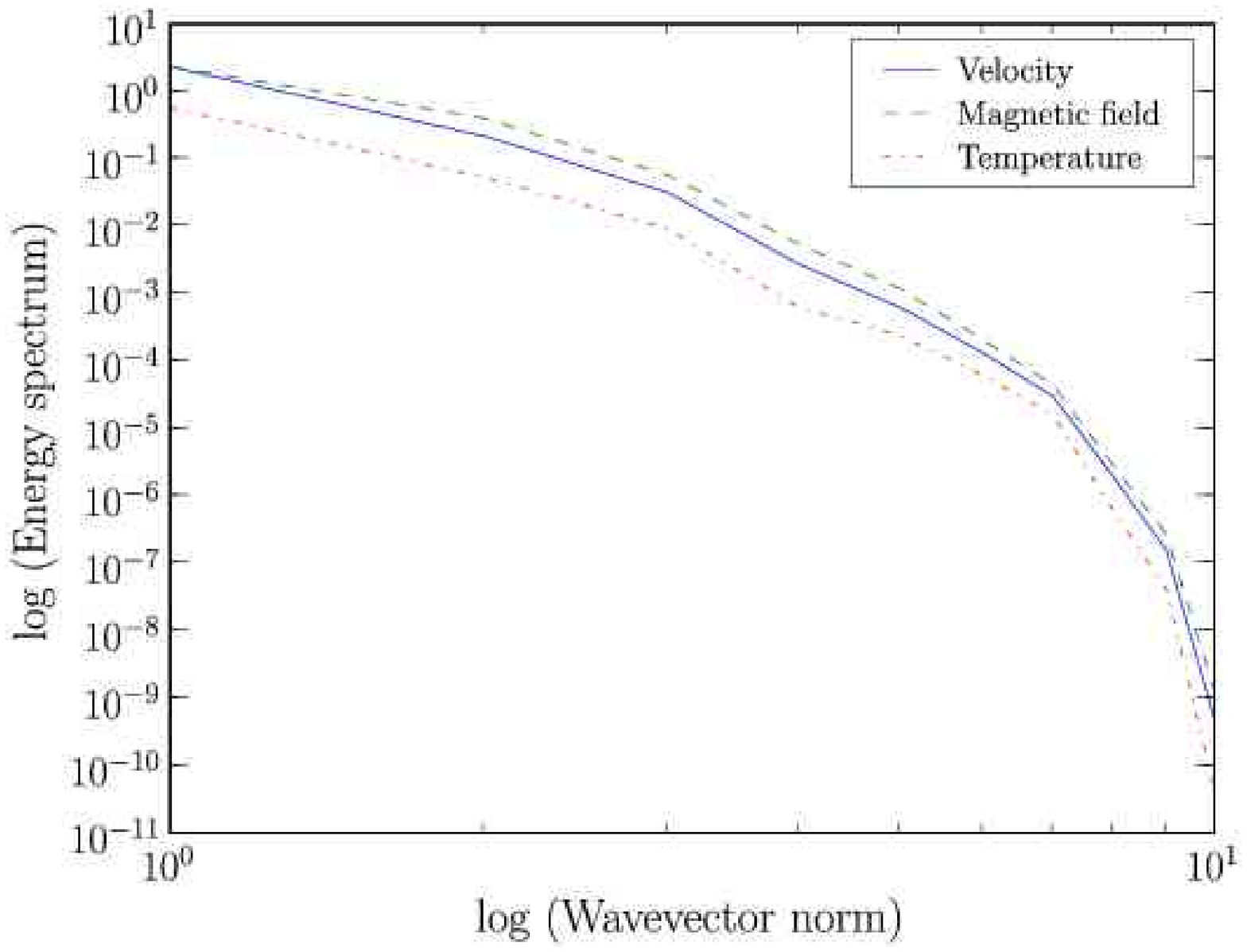}\tabularnewline
$\mathbf{S}_{3}:$ temperature&
$\mathbf{S}_{3}:$ energy spectra\tabularnewline
\end{tabular}\par\end{centering}

\caption{$\mathbf{S}_{3}$ fields and corresponding energy spectra. Anti-symmetry
about the $z\mbox{-axis}$ is observed in all of the displayed fields.\label{fig:S3-fields-and}}
\end{figure}

\begin{figure}[H]
\begin{centering}\begin{tabular}{cc}
\includegraphics[width=6.5cm,keepaspectratio]{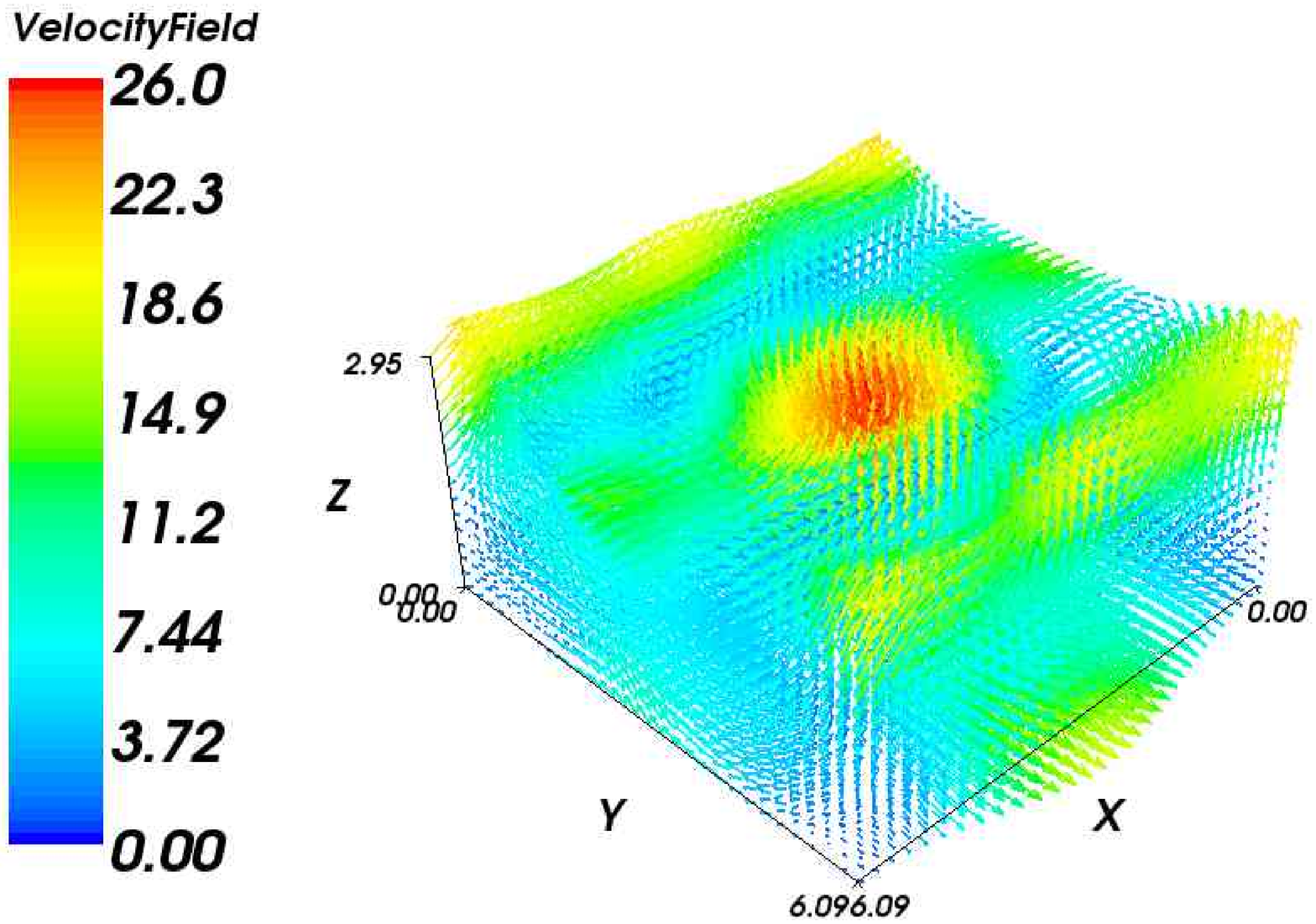}&
\includegraphics[width=6.5cm,keepaspectratio]{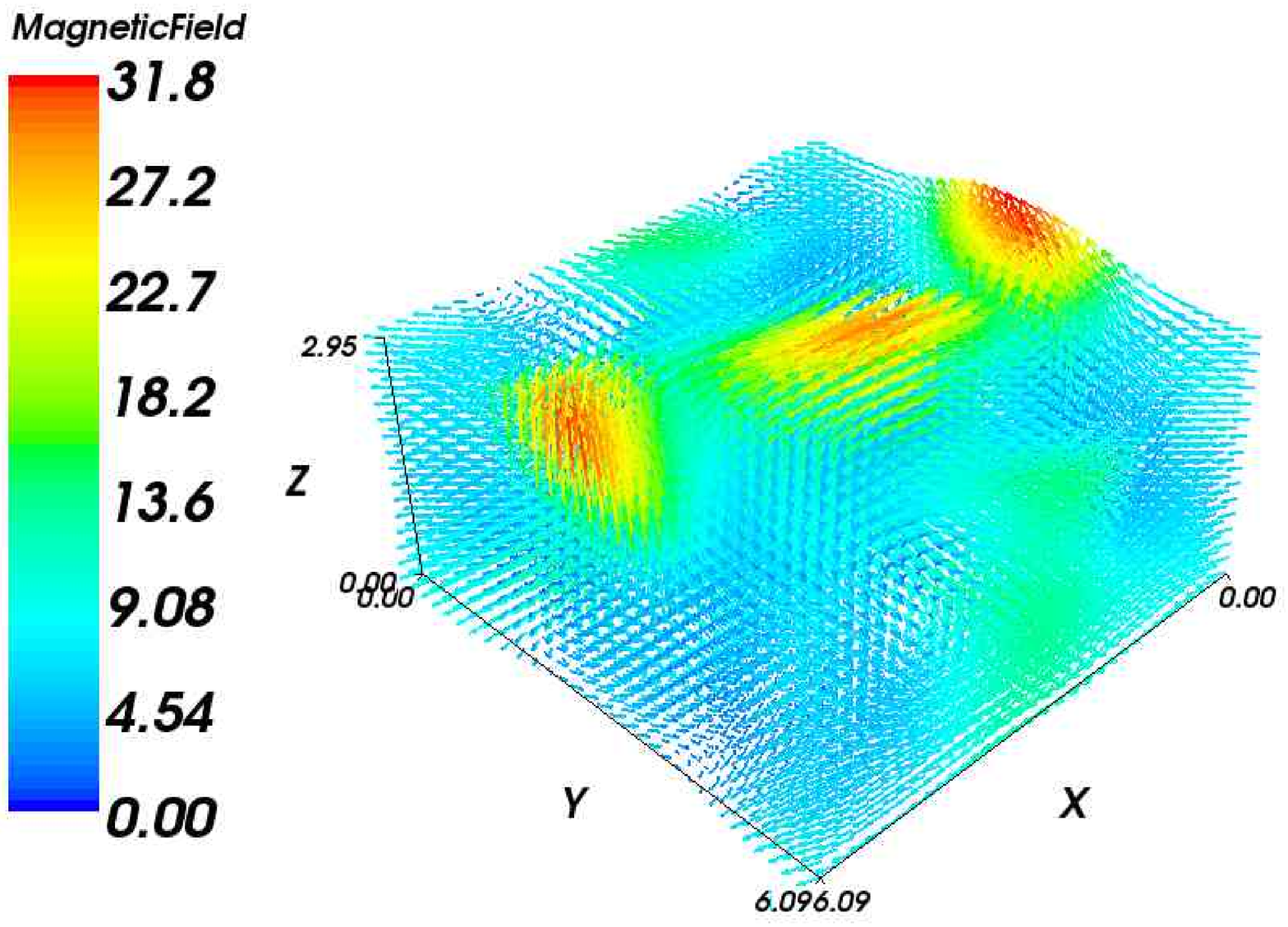}\tabularnewline
$\mathbf{S}_{4}:$ velocity&
$\mathbf{S}_{4}:$ magnetic field\tabularnewline
\includegraphics[width=6.5cm,keepaspectratio]{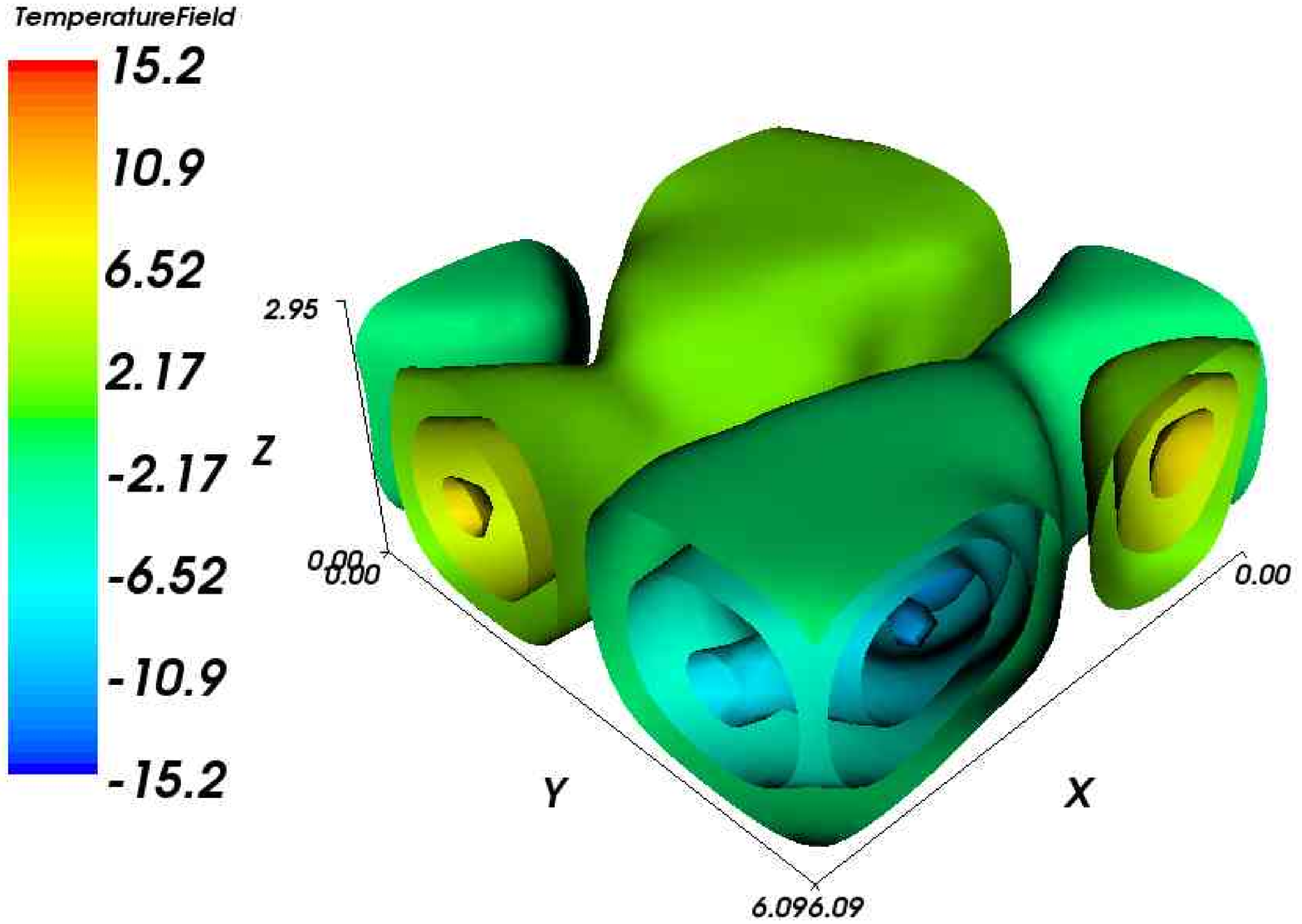}&
\includegraphics[width=6.5cm,keepaspectratio]{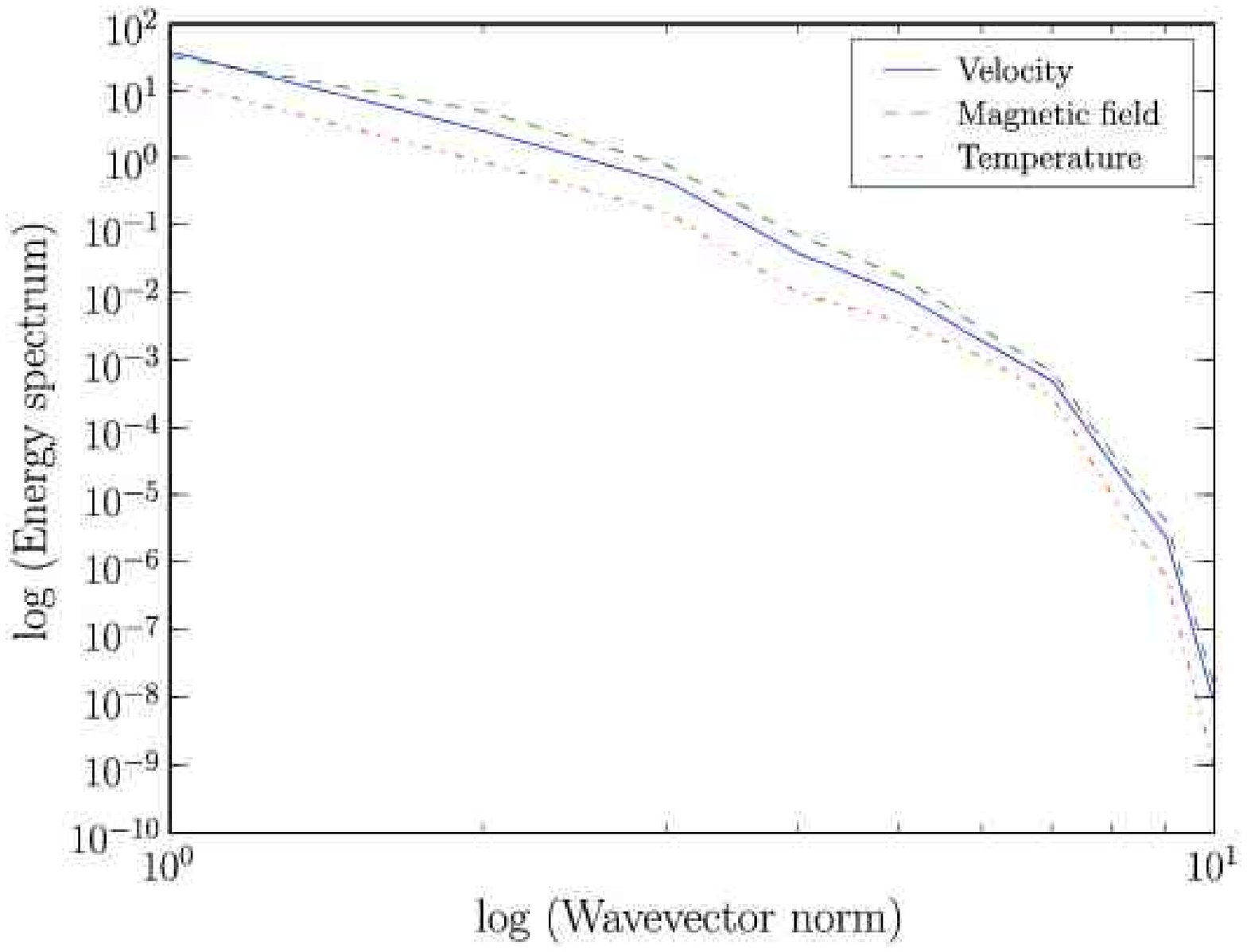}\tabularnewline
$\mathbf{S}_{4}:$ temperature&
$\mathbf{S}_{4}:$ energy spectra\tabularnewline
\end{tabular}\par\end{centering}

\caption{$\mathbf{S}_{4}$ fields and corresponding energy spectra. Anti-symmetry
about the $z\mbox{-axis}$ is observed in all of the displayed fields.\label{fig:S4-fields-and}}
\end{figure}

\begin{figure}[H]
\begin{centering}\begin{tabular}{cc}
\includegraphics[width=6.5cm,keepaspectratio]{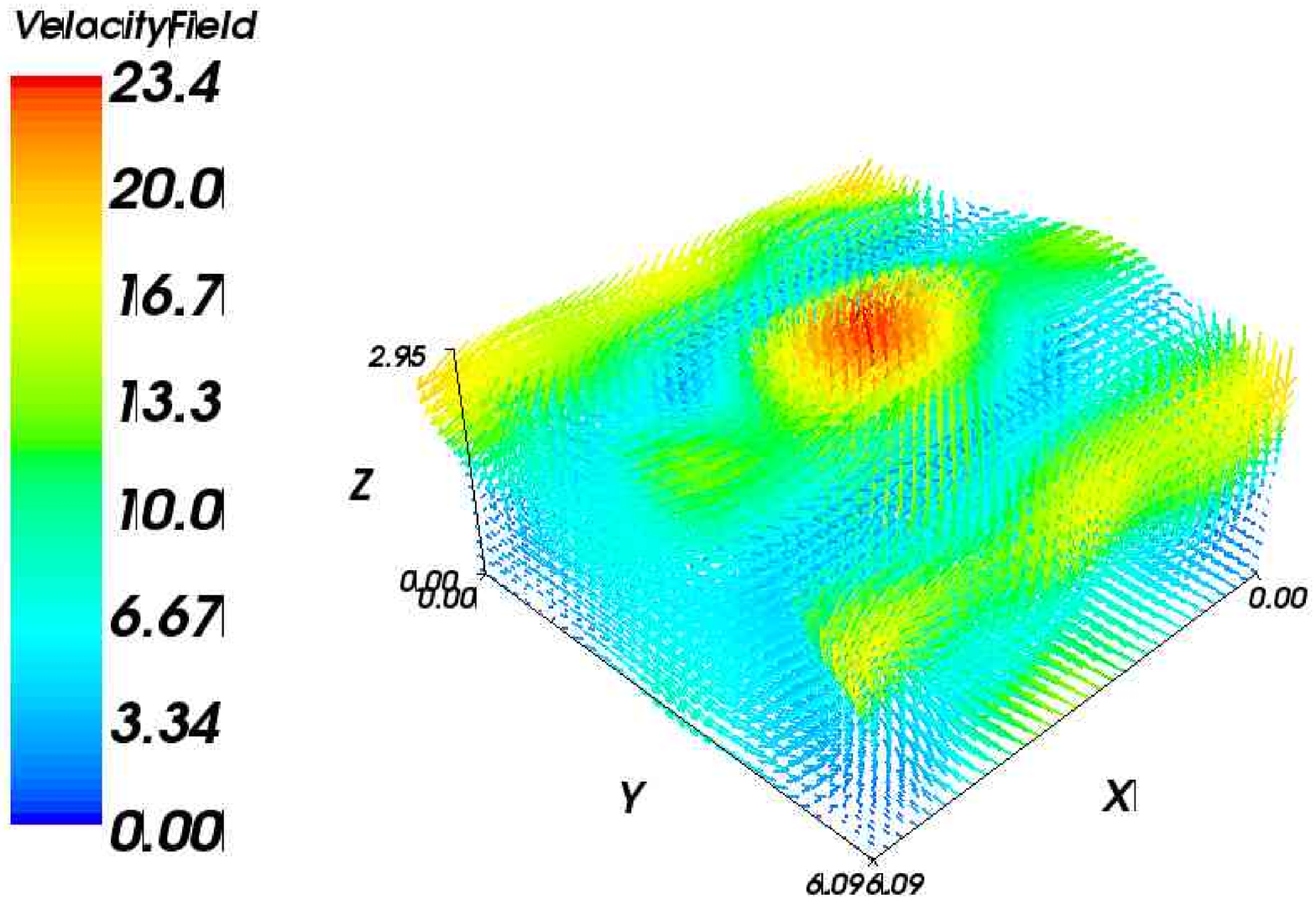}&
\includegraphics[width=6.5cm,keepaspectratio]{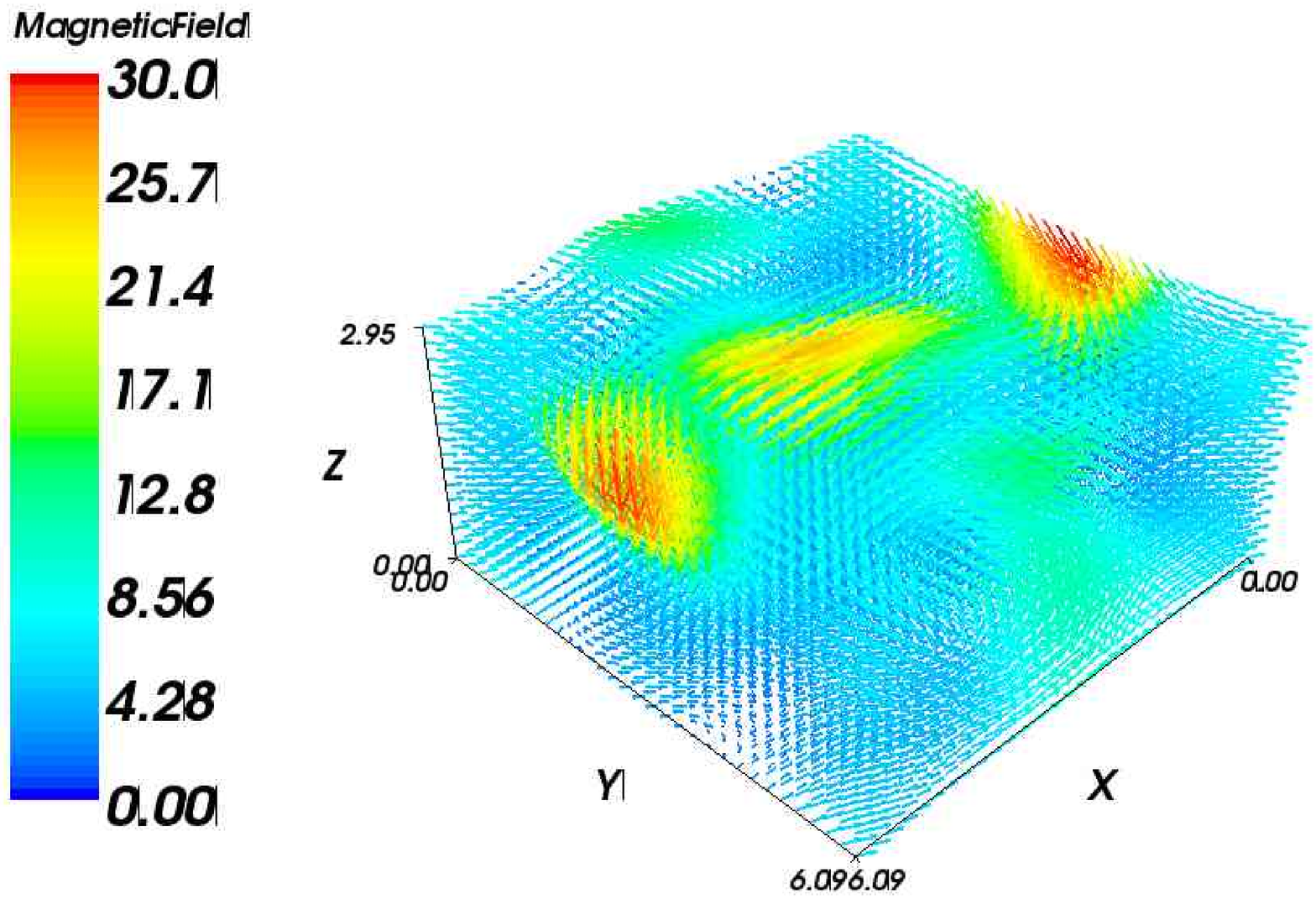}\tabularnewline
$\mathbf{W}^{(0)}:$ velocity&
$\mathbf{W}^{(0)}:$ magnetic field\tabularnewline
\includegraphics[width=6.5cm,keepaspectratio]{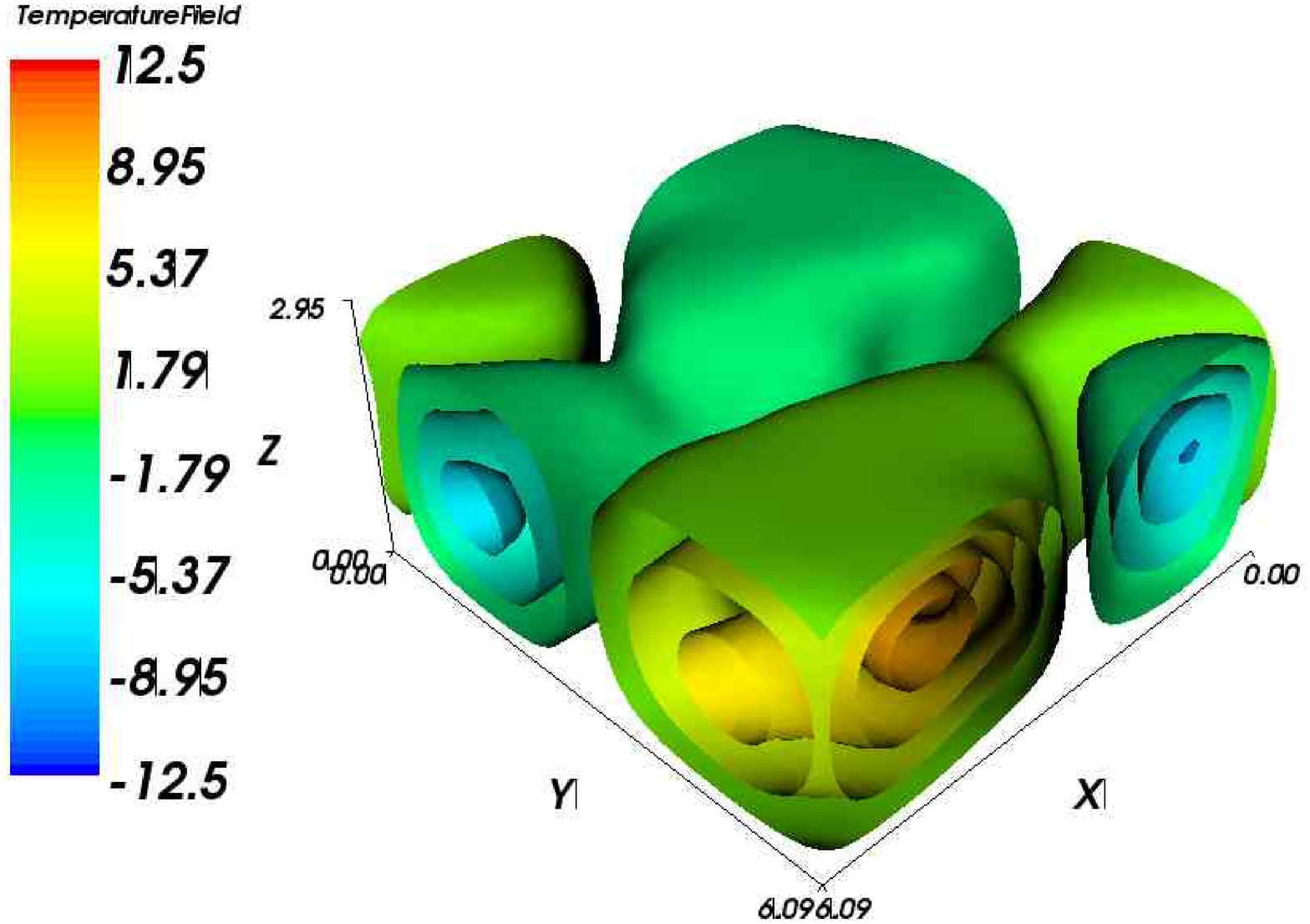}&
\includegraphics[width=6.5cm,keepaspectratio]{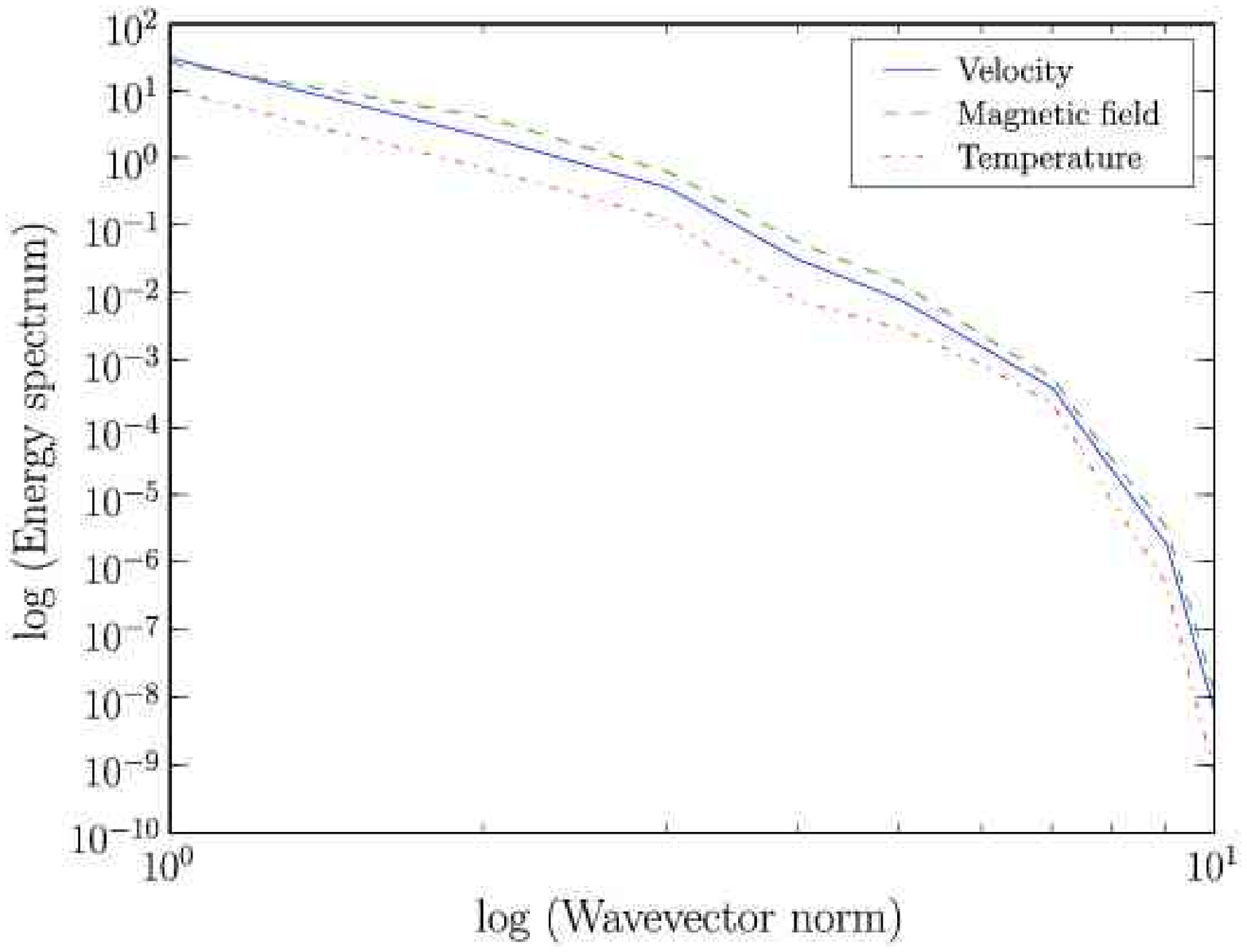}\tabularnewline
$\mathbf{W}^{(0)}:$ temperature&
$\mathbf{W}^{(0)}:$ energy spectra\tabularnewline
\end{tabular}\par\end{centering}

\caption{Short-scale structure of $\mathbf{W}^{(0)}$, for $\lambda_{2}^{max}$.
Anti-symmetry about the $z\mbox{-axis}$ is observed in all of the
displayed fields.\label{fig:W0-fields-and}}
\end{figure}

\begin{figure}[H]
\begin{centering}\begin{tabular}{cc}
\includegraphics[width=6.5cm,keepaspectratio]{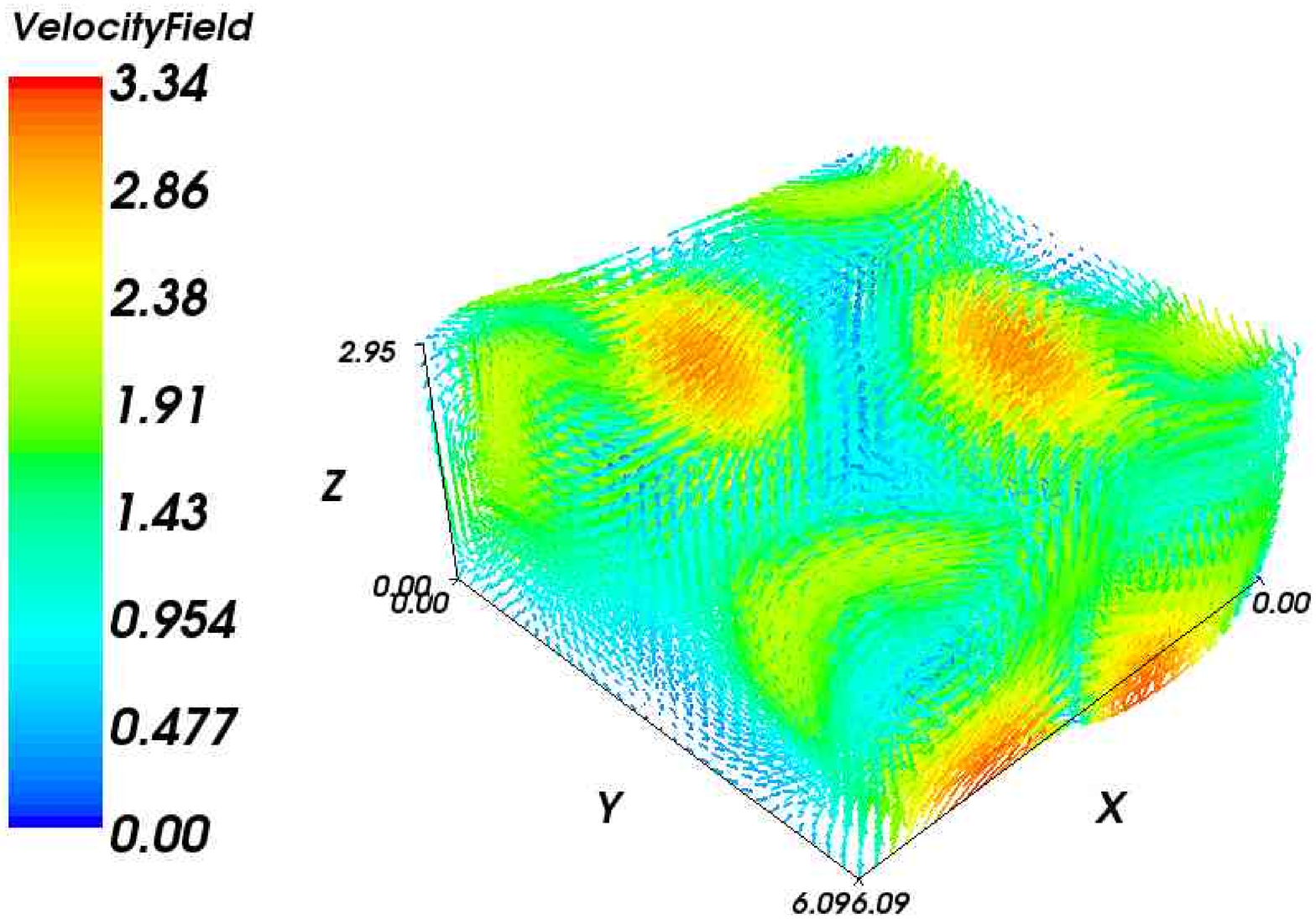}&
\includegraphics[width=6.5cm,keepaspectratio]{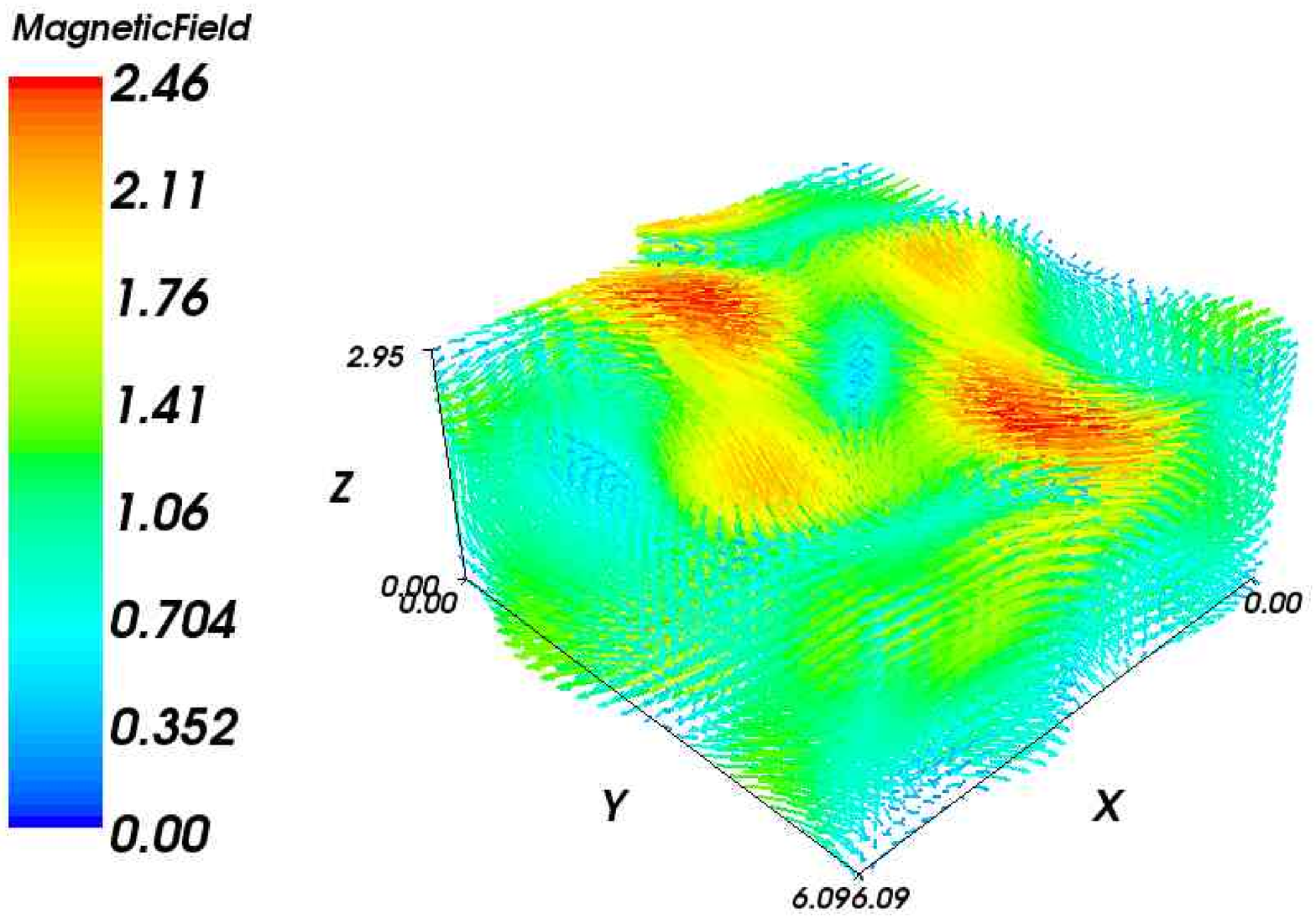}\tabularnewline
$\mathbf{\Gamma}_{11}:$ velocity&
$\mathbf{\Gamma}_{11}:$ magnetic field\tabularnewline
\includegraphics[width=6.5cm,keepaspectratio]{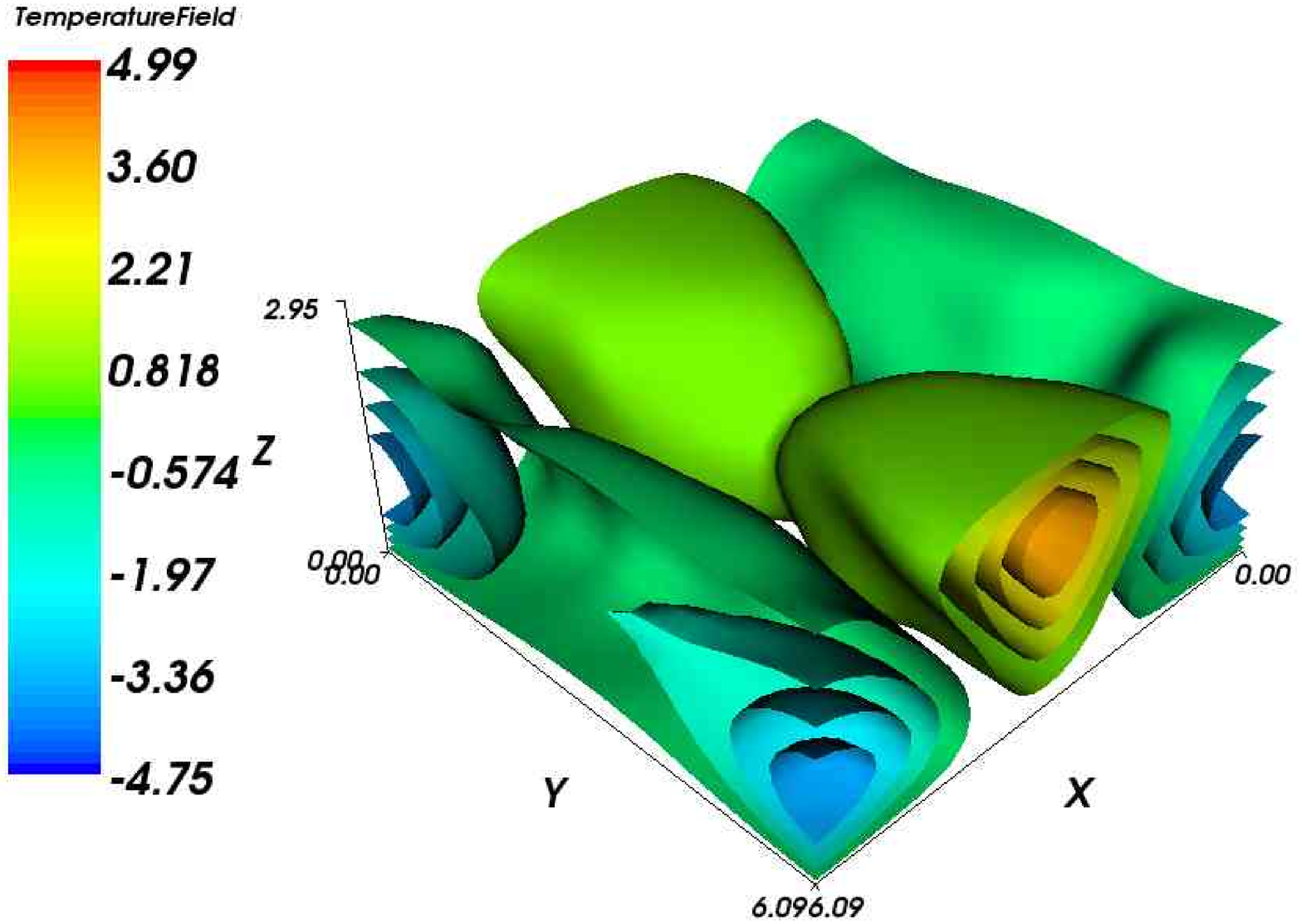}&
\includegraphics[width=6.5cm,keepaspectratio]{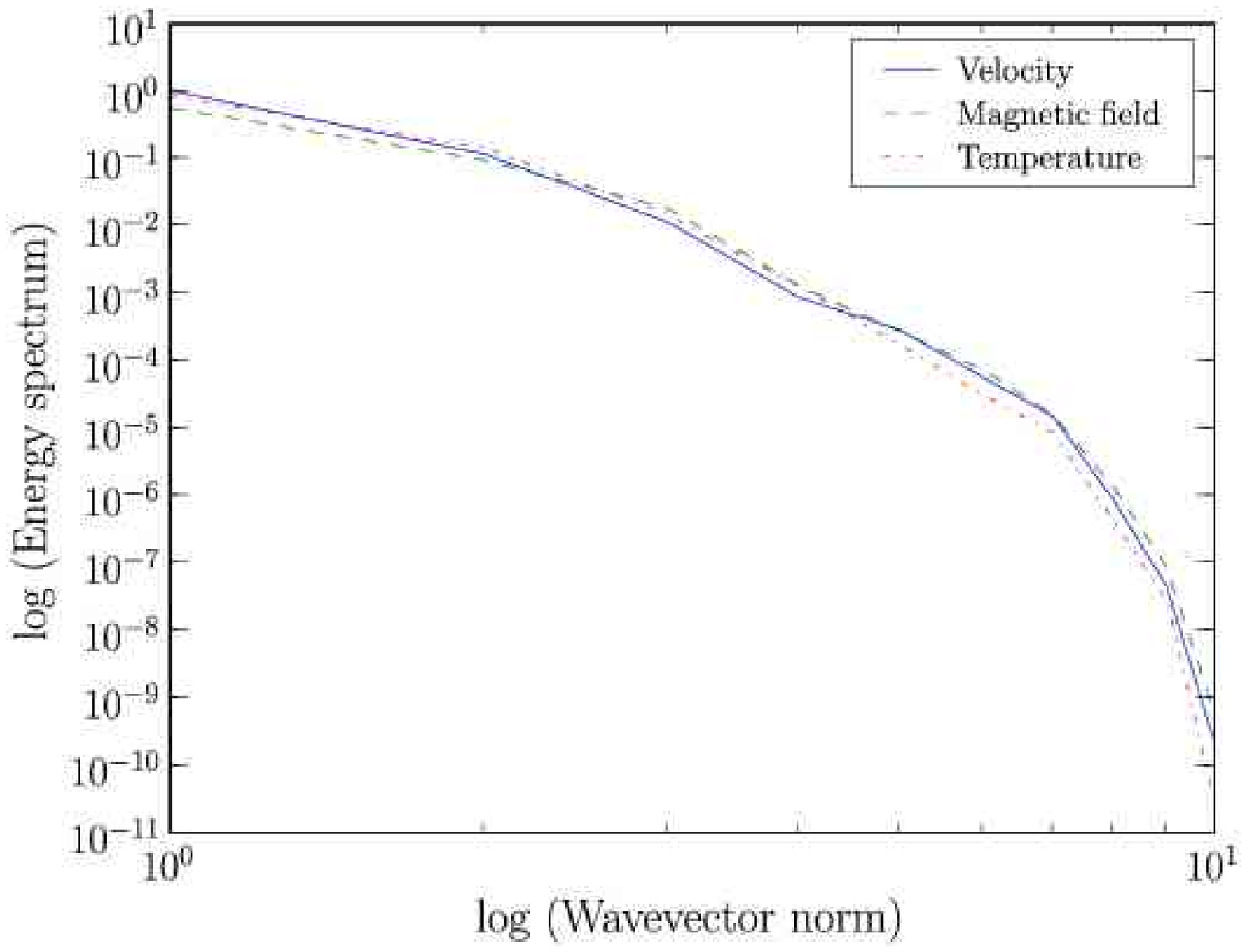}\tabularnewline
$\mathbf{\Gamma}_{11}:$ temperature&
$\mathbf{\Gamma}_{11}:$ energy spectra\tabularnewline
\end{tabular}\par\end{centering}

\caption{$\mathbf{\Gamma}_{11}$ fields and corresponding energy spectra.
Symmetry about the $z\mbox{-axis}$ is observed in all of the displayed
fields.\label{fig:Gamma11-fields-and}}
\end{figure}

\section{Concluding Remarks\label{sec:Concluding-Remarks}}

We have derived an eigenvalue equation for large-scale perturbation
modes of a CHM steady state. On average (over small spatial scales)
the modes are, in the leading order, simple harmonic waves. Their
growth rates are controlled by the combined diffusivity tensor, involving
molecular kinematic viscosity and magnetic diffusivity and an additional
tensor -- the so-called combined eddy (turbulent) diffusivity correction,
which is anisotropic (the entries of the matrix $\mathbf{E}'$ depend
on the direction of the wave vector $\mathbf{q}$), and which intermixes
the influence of the flow and magnetic field. Originally this mutual
influence is due to advection (the influence of the flow on magnetic
field) and the action of the Lorentz force (the influence of the magnetic
field on the flow), but it has now a different algebraic form -- in
particular, unlike in these basic physical laws, it is, on average,
linear.

We have found that about $10\%$ of randomly generated steady CHM
regimes, that are stable to short-scale perturbations, exhibit negative
eddy diffusivity; such steady states are unstable to large-scale perturbations.
However, the growth rate of the perturbation is quadratic in the scale
ratio $\varepsilon$, i.e. it is small. Thus, this instability can
be observed only if the considered CHM steady state is stable to short-scale
perturbations, which would have larger ($O(\varepsilon^{0})$) growth
rates otherwise. Other competing linear instabilities may also persist.
For instance, in thermal convection in a rotating layer with free
boundaries (with no magnetic field present), steady rolls, steady
square cells, standing and travelling waves near the onset were demonstrated
\citep{Po05,Po05mjg} to be unstable to large-scale perturbations
of a particular form (the small angle instability), with the growth
rates scaling as $O(\varepsilon^{2})$ for the rolls and the cells,
and $O(\varepsilon)$ for the waves (here $\varepsilon$ is the smallest
scale ratio in the system); the rolls and square cells possess symmetry
about the vertical axis.

The restriction that the perturbed CHM state is steady can be lifted
\citep{Zhe05fz}. Instead of averaging over fast spatial variables,
averaging over the spatio-temporal domain of fast variables must then
be performed. This allows, in particular, to carry out the stability
analysis of time-periodic CHM states. However, in any case the perturbed
states are required to be symmetric (for instance, symmetry about
the vertical axis, as considered here). Parity invariance is another
type of symmetry consistent with the CHM equations, if the Joule term
is neglected. Like the symmetry about the vertical axis, it guarantees
that no $\alpha$-effect emerges and it allows to construct a second
order combined eddy diffusivity operator by the same method of homogenisation.

Different multiscale expansions are obtained for different sets of
conditions imposed at the horizontal boundaries of the layer. Often
considered alternative boundary conditions include the no-slip condition
for the flow, an isolator outside the fluid layer condition for the
magnetic field, and heat insulating boundaries (zero heat flux condition)
for temperature. They can be of interest, for instance, in geophysical
applications (for which the no-slip condition at the outer kernel
boundaries and the isolator condition at the outer boundary of the
spherical layer are more appropriate than those considered here).
The method of homogenisation that we have used relies on the existence
of constant vector fields in $\textrm{ker}(\mathbf{P}\mathbf{A}^{(0)^{*}}\mathbf{P})$.
In addition to the two constant horizontal vectors in the flow and
magnetic field components considered here, another scalar constant,
a fixed temperature, belongs to $\textrm{ker}(\mathbf{P}\mathbf{A}^{(0)^{*}}\mathbf{P})$,
if Joule heating is neglected ($\sigma=0$) and the zero heat flux
condition for temperature is assumed. Therefore, the procedure of
homogenisation that we have used can be applied if, at least, one
of the following conditions is imposed on the horizontal boundaries:
free boundaries, or conducting boundaries, or no heat flux. Each quantity,
satisfying boundary conditions from this list, increases dimension
of the problem for large-scale mean-fields obtained from the solvability
condition for the equations emerging at order 2. The remaining ones
(temperature in the particular case that we have considered here)
are essentially short-scale and only affect, via solutions of auxiliary
problems at orders 0 and 1, the values of coefficients in the combined
eddy diffusivity tensor for the large-scale quantities. The boundary
conditions for which this approach is not directly applicable (i.e.
the no-slip boundary condition for the flow, or the insulator condition
for the magnetic field, or fixed temperature) can apparently still
be treated by the homogenisation method that we have applied, but
this requires considering boundary layers, increasing significantly
the complexity of the problem.

A common feature of astrophysical convective systems, such as interiors
of planets or stars, is rotation. A straightforward incorporation
of the Coriolis force in the analysis is inconsistent with the homogenisation
procedure that we have used: averaging of the linearised Navier-Stokes
equation (for the free boundary conditions) shows that constant average
non-zero horizontal velocities give rise to a non-zero average Coriolis
force, which can be balanced only by a constant gradient of pressure.
This suggests an unbounded linear growth of pressure, which is not,
however, unphysical: in a rotating system only pressure can offset
the centrifugal force. The simplest way to overcome the resultant
algebraic difficulties is to consider the vorticity equation, for
which the methods for construction of the two-scale expansion are
applicable without any modifications.

\subsection*{Acknowledgements}

We have benefited from stimulating discussions with U. Frisch, A.M.
Soward and K. Zhang. MB was financially supported by FCT (Portugal,
studentship BD/8453/2002). VZ is grateful to the Royal Society, the
French Ministry of Education, CMAUP (Portugal) and the Russian Foundation
for Basic Research (grant 04-05-64699).

\bibliographystyle{epj}

\bibliography{mhdconv_epjb_bib}

\end{document}